\documentclass[%
 aip,
 pop,
 amsmath,amssymb,
 preprint,%
]{revtex4-1}
\usepackage{graphicx}
\usepackage{xcolor}
\usepackage{caption}
\usepackage{subcaption}
\usepackage{multirow}

\newcommand{\rev}[1]{{#1}}
\newcommand{\rrev}[1]{{#1}}

\draft 

\DeclareUnicodeCharacter{2212}{-}

\begin{document}


\title{Latent Space Mapping: Revolutionizing Predictive Models for Divertor Plasma Detachment Control}



\author{Ben Zhu}
\email[]{zhu12@llnl.gov}
\affiliation{Lawrence Livermore National Laboratory, Livermore, CA 94550, USA}
\author{Menglong Zhao}
\affiliation{Lawrence Livermore National Laboratory, Livermore, CA 94550, USA}
\author{Xue-Qiao Xu}
\affiliation{Lawrence Livermore National Laboratory, Livermore, CA 94550, USA}
\author{Anchal Gupta}
\affiliation{Oak Ridge Associated Universities, Oak Ridge, TN 37830, USA}
\author{KyuBeen Kwon}
\affiliation{Oak Ridge Associated Universities, Oak Ridge, TN 37830, USA}
\author{Xinxing Ma}
\affiliation{General Atomics, San Diego, CA 92186, USA}
\author{David Eldon}
\affiliation{General Atomics, San Diego, CA 92186, USA}



\date{5 June 2025}

\begin{abstract}

The inherent complexity of boundary plasma, characterized by multi-scale and multi-physics challenges, has historically restricted high-fidelity simulations to scientific research due to their intensive computational demands. Consequently, routine applications such as discharge control and scenario development have relied on faster but less accurate empirical methods. This work introduces DivControlNN, a novel machine-learning-based surrogate model designed to address these limitations by enabling quasi-real-time predictions (i.e., $\sim0.2$ ms) of boundary and divertor plasma behavior. Trained on over 70,000 2D UEDGE simulations from KSTAR tokamak equilibria, DivControlNN employs latent space mapping to efficiently represent complex divertor plasma states, achieving a computational speed-up of over $10^8$ compared to traditional simulations while maintaining a relative error below 20\% for key plasma property predictions. During the 2024 KSTAR experimental campaign, a prototype detachment control system powered by DivControlNN successfully demonstrated detachment control on its first attempt, even for a new tungsten divertor configuration and without any fine-tuning. These results highlight the transformative potential of DivControlNN in overcoming diagnostic challenges in future fusion reactors by providing fast, robust, and reliable predictions for advanced integrated control systems. 

\end{abstract}


\maketitle 

\section{Introduction}\label{sec:intro}

Heat and particle exhaust represents a significant \rev{Research and Development (R\&D)} challenge in magnetic fusion energy research. In future fusion power plants, fusion-generated power must be effectively exhausted to prevent reactor wall damage. For instance, many fusion reactor designs rely on deuterium-tritium (D-T) fusion due to its relatively large reaction cross-section at achievable temperatures. In this reaction, approximately one-fifth of the fusion energy is carried by the 3.5 MeV alpha particles, which are expected to collide with and further heat deuterium and tritium particles. In other words, a substantial amount of fusion power, e.g., roughly 100 MW for a 500 MW DT-based fusion reactor, will convert into the plasma thermal energy and be deposited to a narrow region, e.g., on the order of millimeters on the divertor target plates in toroidal devices.~\cite{eich2013scaling} Without mitigation, this level of heat load exceeds the limits of the most advanced materials by orders of magnitude; therefore, a large portion of the heat must be radiatively dissipated before it reaches the divertor target plates and plasma-facing components. Additionally, the exhaust plasma needs to enter the so-called detached state, either partially or fully, to prevent excessive plasma-material interactions. In the detached state, divertor plasma is cooled away from the divertor targets by impurity radiation and high neutral gas density. This phenomenon is termed \textit{detachment} because most of the plasma recombines into neutral gas before reaching the divertor plate; hence, the plasma is ``detached" from the plasma-facing components. 
Divertor plasma detachment is essential for maintaining low target plate temperature and reducing ion flux to the divertor, thereby minimizing sputtering and limiting the total heat flux to prevent divertor melting. For example, in tungsten divertors, the plasma temperature at the target plates must drop to the eV range (i.e., the sputtering limit~\cite{brezinsek2019erosion}), and the divertor heat flux must be reduced to below 10 MW/m$^2$ (i.e., the melting limit~\cite{pitts2019physics}). 
However, seeding impurities and puffing neutral gas in the divertor region often degrade core plasma performance~\cite{stangeby2018basic}.

Thus, identifying an optimal operational space, establishing and maintaining an appropriate degree of plasma detachment, and ensuring a stable and well-positioned ``detachment front" are non-trivial requirements, and real-time detachment control is an active area of research. For example, real-time feedback detachment controls have been developed and tested in various experiments, such as DIII-D/EAST~\cite{wang2022achievements}, KSTAR~\cite{eldon2022enhancement}, and TCV~\cite{ravensbergen2021real}. These control systems rely on in-situ diagnostic measurements, such as the degree of detachment control based on ion saturation current measurements on the divertor plates from Langmuir probes~\cite{eldon2022enhancement}, or peak radiation location control using C-III emission images from the MANTIS diagnostic~\cite{ravensbergen2021real}. Unfortunately, these diagnostic tools may not be feasible in future reactor-grade machines due to spatial constraints and/or the radiative environment. In addition, the prolonged latency of actuators in future reactors presents further challenges in designing an effective feedback control system. Therefore, a more sophisticated controller, such as a model-based controller that does not solely rely on in-situ diagnostics but incorporates an embedded physics model, would be extremely beneficial.

Historically, due to the inherent multi-scale and multi-physics nature of divertor plasma~\cite{krasheninnikov2017physics}, axisymmetric boundary transport models for tokamaks, such as UEDGE~\cite{rognlien1999two} and SOLPS~\cite{schneider2006plasma,wiesen2015new}, are computationally time-consuming. For instance, even low-fidelity, one-dimensional simulations, e.g., SD1D~\cite{dudson2019role}, take several minutes, while higher-fidelity, two-dimensional runs could take from hours to days to complete, limiting their applications to physics research. Furthermore, these models are sensitive to initial conditions, and a converged solution is not always guaranteed. However, emerging machine learning techniques offer an alternative solution to the detachment control challenge by constructing fast yet accurate surrogate models through a data-driven approach.

Recently, several machine learning-based surrogate models for divertor plasmas have been developed. For instance, a surrogate model trained on approximately 170,000 one-dimensional UEDGE simulation data has been constructed to predict steady-state boundary plasma solutions containing both upstream and downstream plasma conditions, as well as the poloidal radiation distribution, based on the global discharge parameters for the DIII-D tokamak~\cite{zhu2022data}. Even without any optimization and code acceleration, this model is four orders of magnitude faster than direct simulations, e.g., 30 ms vs a few minutes.
Similarly, machine learning emulators have been explored for the MAST-U tokamak~\cite{holt2024tokamak} using thousands of one-dimensional Hermes-3 simulations~\cite{dudson2024hermes} with forward neural networks, attaining solutions in just 1 ms. 
\rev{Moreover, neural networks have also been employed to predict the electron temperature at the outer divertor target, based on a database of thousands of two-dimensional SOLPS simulations with eight varying tokamak parameters~\cite{dasbach2023towards}.}
Beyond steady-state solution prediction, neural partial differential equation (PDE) surrogates have been explored to predict the dynamical evolution of divertor plasma~\cite{poels2023fast} using hundreds of DIV1D simulations~\cite{derks2022benchmark}, achieving sub-real-time speed performance. 
\rev{Reference~\cite{wiesen2024data} offers a concise yet comprehensive review of machine learning applications in modeling plasma exhaust processes for fusion reactors.}

\begin{figure}[h]
    \centering
    \includegraphics[width=\linewidth]{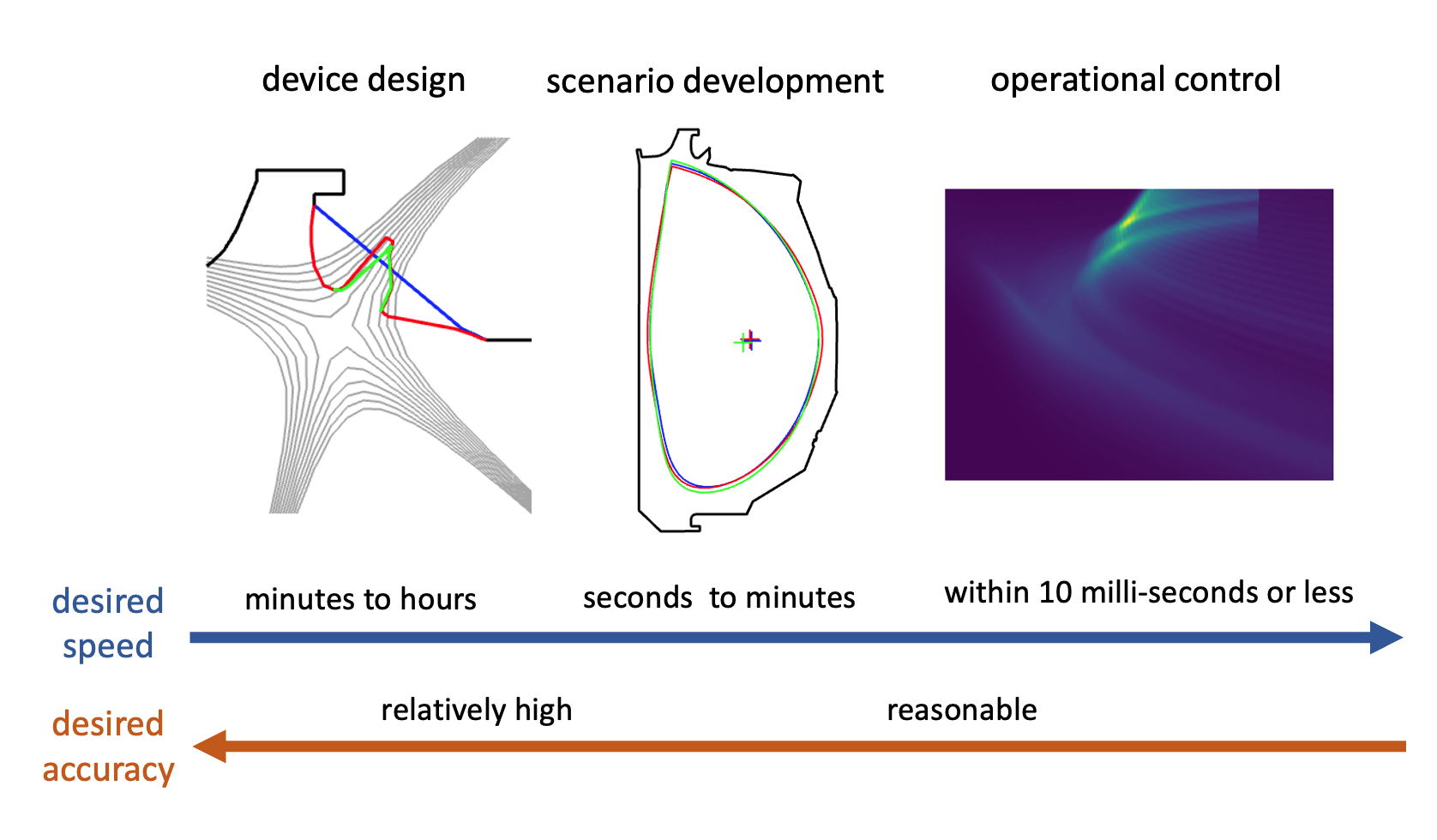}
    \caption{\rev{Illustrative examples of boundary and divertor modeling applications along with their performance requirements. From left to right: the evolution of the DIII-D upper divertor configuration, transitioning from open (blue line) to Small Angle Slot (SAS, red line) and to SAS V-shaped (green line) configurations; reconstructed last closed flux surfaces and plasma centers of a DIII-D discharge at various time points; and a synthetic divertor emissivity image for a DIII-D discharge generated using UEDGE simulations.}}
    \label{fig:apps}
\end{figure}

In this work, we have extended our previous proof-of-principle study~\cite{zhu2022data} to incorporate two-dimensional UEDGE simulation results of KSTAR tokamak, thereby broadening the scope to more realistic applications, such as detachment control. The resulting surrogate model, named DivControlNN, provides quasi-real-time predictions (approximately $0.1$ ms) of critical divertor and outboard midplane plasma information based on real-time discharge and diagnostic measurements, utilizing the latent space mapping technique. It is worth mentioning that although the primary focus of this paper is on detachment control, the application of our approach is more general, as illustrated in Figure~\ref{fig:apps}. Different surrogate models tailored to various applications, such as device design and discharge scenario development, can be constructed and trained to meet the required speed and accuracy specifications. 

The rest of the paper is organized as follows. Section~\ref{sec:method} introduces the data-driven surrogate model approach and the two-stage training process employed in this study. Section~\ref{sec:prep} describes the development of a surrogate model for detachment control experiments on the KSTAR tokamak, along with the generation of 2D UEDGE simulation data for model training and validation. In Section~\ref{sec:train}, the performance of individual neural networks is evaluated, while Section~\ref{sec:perf} provides a comprehensive performance analysis of the surrogate model, including its prediction accuracy, model uncertainty, speed, and comparisons with direct UEDGE simulation results. Section~\ref{sec:app} presents preliminary results from KSTAR experiments on detachment control using a prototype controller based on the surrogate model. Finally, Section~\ref{sec:sum} summarizes the key findings of our study.

\section{Latent space mapping}\label{sec:method}

Various methods exist to build surrogate models that can predict solutions without performing actual simulations. The method used in this study leverages the concept of latent space representation (LSR). This approach involves first identifying an LSR description of the system alongside global parameters (e.g., model inputs for simulations or control variables in experiments) and detailed diagnostic measurements. Next, a projection using neural networks (NNs) is established between these three different descriptions, utilizing the latent space as an intermediate step to predict diagnostic measurements from the global parameters.

This idea was initially explored in inertial confinement fusion research (i.e., unmagnetized plasmas)~\cite{anirudh2020improved}, where self-consistent image diagnostics and scalar quantities were predicted in response to various input parameters (e.g., target and laser settings). Subsequently, a proof-of-principle study using simplified 1D UEDGE simulations was conducted to explore this method for magnetized plasmas, successfully demonstrating that complex divertor plasma states can be efficiently represented in a reduced, low-dimensional latent space~\cite{zhu2022data}.

\begin{figure}[h]
    \centering
    \includegraphics[width=\linewidth]{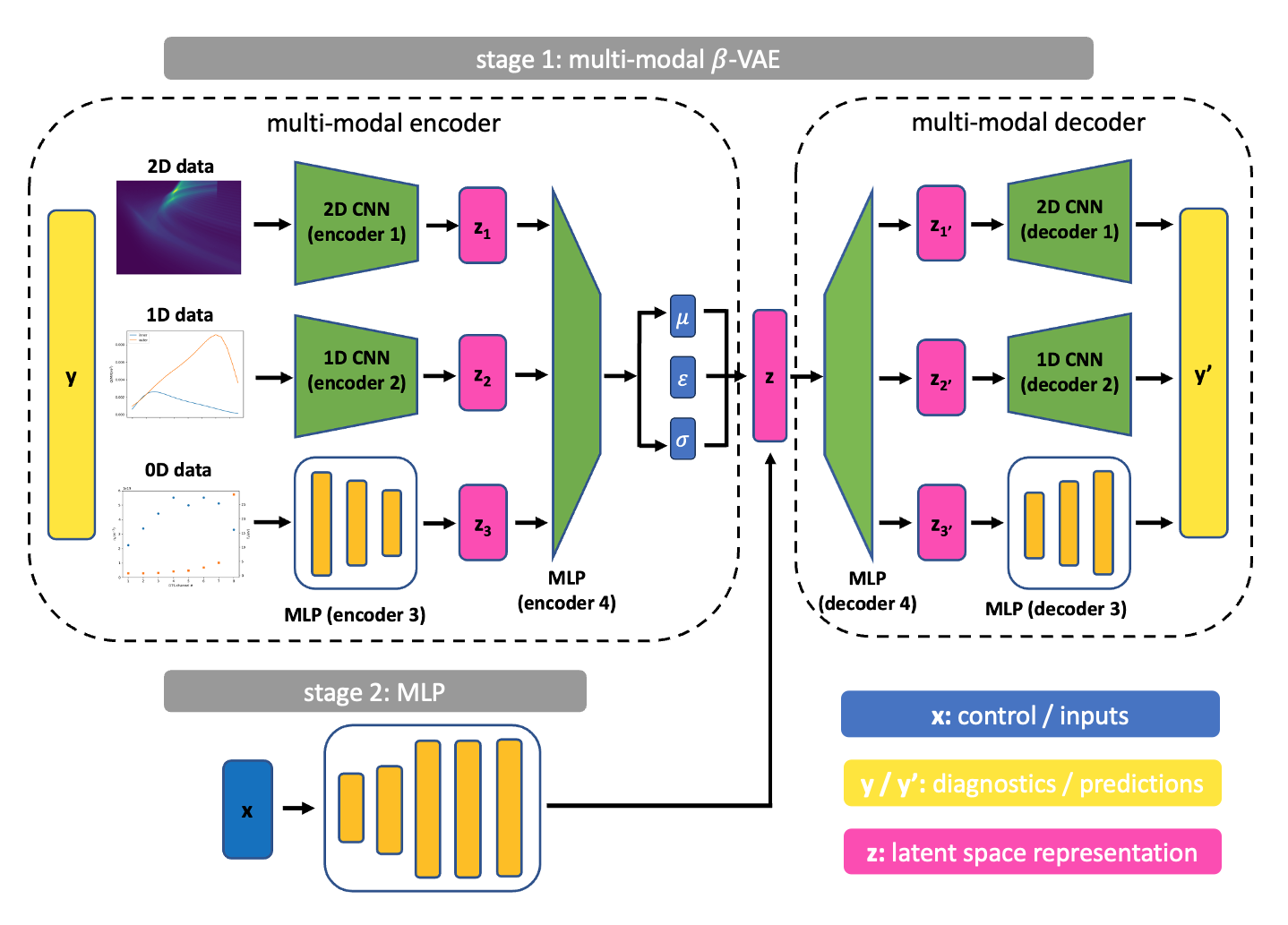}
    \caption{Illustration of the two-stage surrogate model training process. In the first stage, a multi-modal $\beta$-variational autoencoder ($\beta$-VAE) is employed to identify the corresponding latent space representation (LSR, denoted as $\mathbf{z}$) for a set of plasma diagnostics ($\mathbf{y}$). In the second stage, a multi-layer perceptron (MLP) is trained to map discharge control parameters ($\mathbf{x}$) to the reduced-dimension LSR. The complete surrogate model thus consists of the forward LSR prediction neural network ($\mathbf{x}\to \mathbf{z}$) and the multi-modal decoder ($\mathbf{z}\to \mathbf{y'}$).}
    \label{fig:nns}
\end{figure}

A similar two-stage training process is employed to construct fast and robust surrogate models for the 2D boundary plasma systems, as illustrated in Figure~\ref{fig:nns}. The first stage identifies an appropriate latent space representation (LSR) of plasma states by compressing the desired synthetic diagnostics through a multi-modal beta-variational autoencoder ($\beta$-VAE~\cite{higgins2017beta}). In the multi-modal $\beta$-VAE, diagnostics are organized into different groups according to their modality before being compressed. The second stage projects a set of discharge and control parameters to their corresponding LSR using a multi-layer perceptron (MLP). By combining the trained MLP with the decoder network from the autoencoder, this new data-driven surrogate model can predict a consistent set of plasma conditions based on a limited number of discharge and control parameters.

In comparison to the previous study, a significant improvement is the extension of the training data and the model from a one-dimensional flux tube near the separatrix to the entire two-dimensional poloidal plane. While the one-dimensional model captures essential divertor plasma physics and can qualitatively match experimental measurements by adjusting empirical dissipation coefficients~\cite{dudson2019role,body2024detachment}, it has several limitations that hinder its application for detachment control in real experiments. First, the one-dimensional model lacks perpendicular diffusion dynamics, which restricts the radial spreading of particle and heat loads. This often leads to over-predictions of radiation and plasma quantities at the target. Additionally, it presents a different temperature cliff threshold and underlying physics compared to experimental observations. Furthermore, the selection of a specific flux tube (often the one in the SOL next to the separatrix) inherently limits the model's applicability for controlling the entire divertor region. This limitation is particularly relevant in scenarios such as partial detachment or peak radiation locations, which may not be encompassed within this particular flux tube. Recognizing that a data-driven surrogate model is only as effective as the underlying model that generates the training data, it is essential to extend our work to a two-dimensional axisymmetric transport model for practical applications.

This study focuses on the prediction of steady-state solutions. We recognize that divertor plasma, being a highly nonlinear system, can exhibit bifurcation and hysteresis; therefore, a time-dependent prediction of plasma state trajectories is preferred. However, our model remains valid for the following reasons: the hysteresis window is often very narrow, and the prolonged latency (e.g., from actuator response) could exceed the plasma relaxation time in future reactors, indicating temporal scale separation. That being said, building a reduced-order model (ROM) of divertor plasma state evolution adopting the same latent space mapping technique is feasible and will be the focus of our future study.



\section{Design of surrogate model}\label{sec:prep}

In this section, we outline the preparatory work undertaken prior to surrogate model training. This involves the selection of key model inputs and outputs, carefully tailored for real-time detachment control in tokamak experiments such as KSTAR, with a particular focus on ensuring compatibility and seamless integration with the existing Plasma Control System (PCS). Furthermore, we detail the process of generating the UEDGE database utilized for surrogate model development.

\subsection{Model inputs and outputs}\label{subsec:modelio}
\begin{figure}[h]
    \centering
    \includegraphics[width=\linewidth]{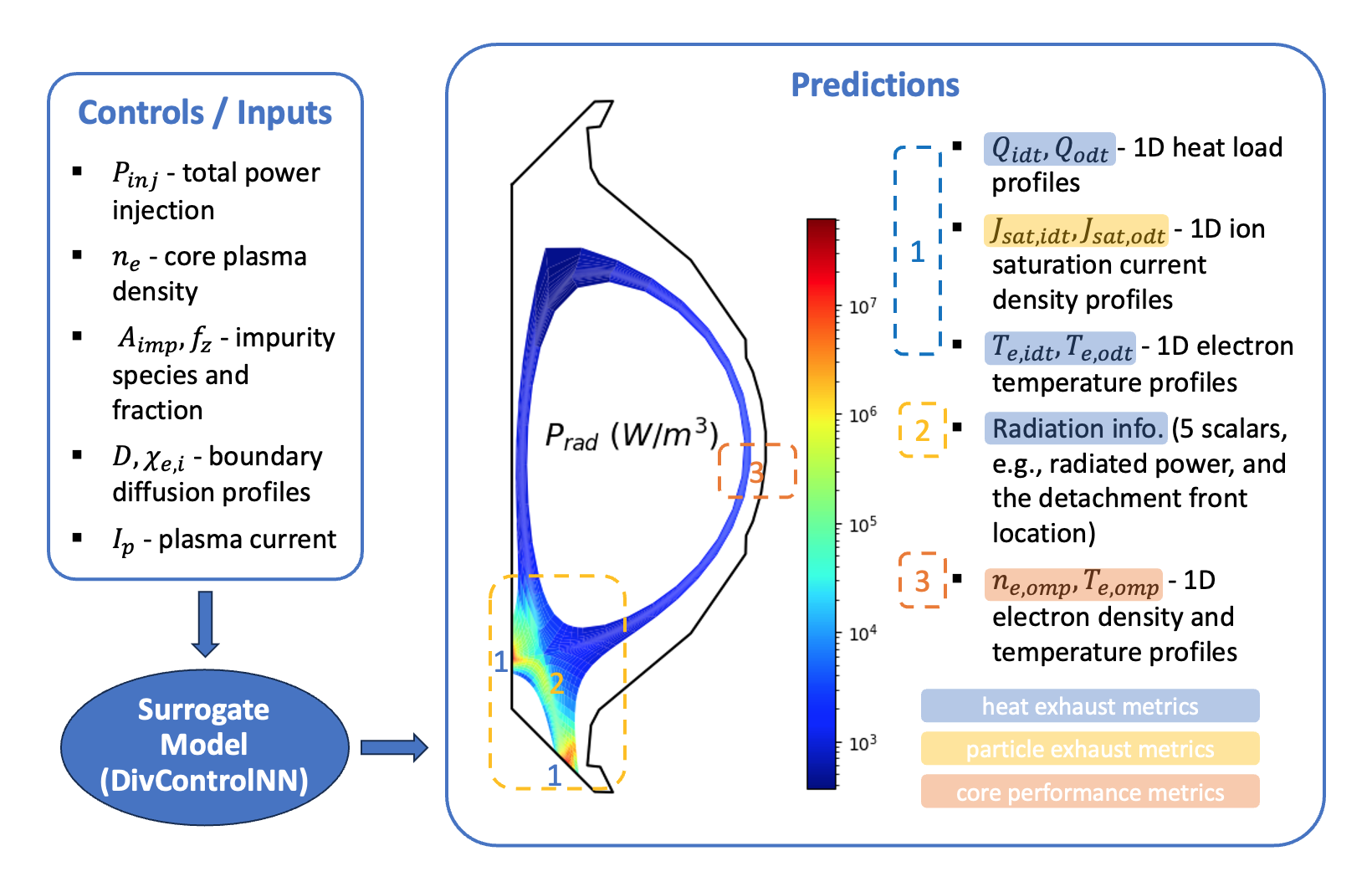}
    \caption{The inputs and outputs of DivControlNN tailored for KSTAR detachment control experiments. The inputs consist of real-time discharge control parameters (e.g., $P_\mathrm{inj},I_p$), diagnostic data (e.g., $n_\mathrm{e,core}$), and estimates such as diffusivity and impurity fraction. This model is capable of predicting plasma conditions at multiple locations, including the divertor target plates (denoted as location 1), divertor region (location 2), and the outboard mid-plane (location 3). These predictions can be effectively utilized for managing heat and particle exhaust and optimizing fusion performance.}
    \label{fig:io_kstar}
\end{figure}

The surrogate model developed here in this work, DivControlNN, is specifically designed for real-time detachment control and is intended for integration into the Plasma Control System (PCS) of tokamak experiments such as KSTAR. Consequently, the inputs and outputs of the surrogate model have been carefully selected to fulfill this purpose. Figure~\ref{fig:io_kstar} illustrates the inputs and outputs of the surrogate model.

The model inputs include the total power injected into the boundary region $P_\mathrm{inj}$, core electron density $n_\mathrm{e,core}$ (near $\psi_n=0.85$), impurity species $A_\mathrm{imp}$ and its fraction $f_Z$, effective boundary diffusion profiles $D$ and $\chi_{e,i}$, as well as the discharge plasma current $I_p$. These variables are critical discharge and control parameters that influence boundary and divertor plasma conditions and can be obtained through real-time control (e.g., $P_\mathrm{inj}$ and $I_p$), diagnostics (e.g., $n_\mathrm{e,core}$ from interferometry), or estimation (i.e., impurity fraction and diffusion coefficients). 
It is important to note that while the current study uses impurity fraction as the model input due to the fixed-fraction impurity model used in the UEDGE simulation data generation process, future stages will implement a multi-charge-state impurity model, updating this input to the impurity seed rate for improved alignment with experimental control.

The model outputs encompass plasma conditions at multiple locations, including (1) electron temperature, ion saturation current, and heat load profiles at both inner and outer divertor plates; (2) radiation information (e.g., overall and divertor radiation fraction, peak radiation power, and detachment front location); and (3) outboard midplane electron density and temperature profiles. For the radiation in the divertor region, an early attempt to predict the entire 2D radiation distribution yielded promising results; however, this approach resulted in significantly slower performance. Therefore, for real-time control purposes, the divertor radiation information is extracted and characterized using a few scalar values. Nevertheless, these outputs provide a more comprehensive description of boundary plasma than the actual experimental diagnostics, such as inner divertor measurements. Additionally, they cover divertor particle exhaust and upstream plasma performance predictions, making the model easily extendable for multi-objective control development in the future.


\subsection{KSTAR database generation}\label{subsec:database}

Starting with the equilibrium of KSTAR discharge \#22849 at 5600 ms, the UEDGE code is set up to obtain steady-state boundary plasma solutions with randomly sampled inputs. These inputs include varying injection power, upstream density, effective diffusion, impurity species (e.g., carbon, nitrogen, neon) and their fractions, plasma current (and consequently, divertor leg connection length), with and without drift effects, among other parameters. Throughout this study, more than 150,000 converged simulations with various settings were obtained, covering a large portion of the parameter range in the KSTAR operational space.

\begin{figure}[h]
    \centering
    \includegraphics[width=0.48\linewidth]{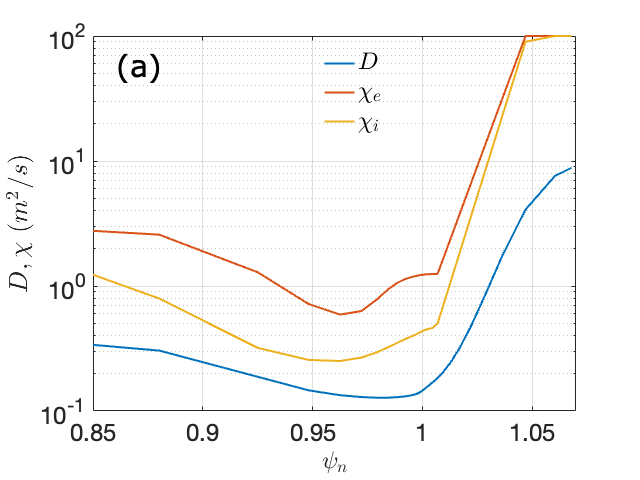}
    \includegraphics[width=0.5\linewidth]{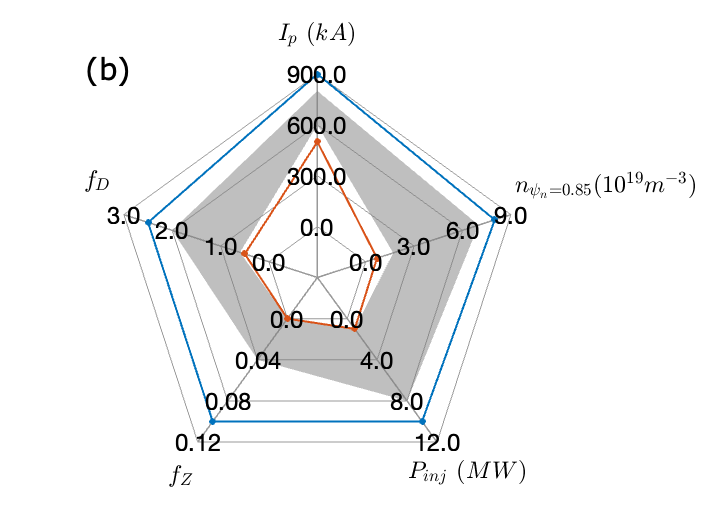}
    \caption{(a) Reference effective diffusion coefficient profiles used in UEDGE simulations, and (b) input parameter range of UEDGE simulations. The red and blue lines represent the minimum and maximum values simulated for KSTAR within the project, while the shaded area indicates the parameter space covered for the study reported here in this paper.}
    \label{fig:kstar_para_space}
\end{figure}

The surrogate model presented in this paper has been trained and tested on a subset of the entire database, comprising approximately 87,000 2D UEDGE KSTAR simulations with carbon as the impurity and with full drift effects (favorable $B_t$). A few assumptions were made to simplify data generation and reduce computational costs:
(1) the shape of effective diffusion coefficient profiles is fixed, allowing the use of a single parameter $f_D$ to scale the reference profiles within the separatrix and near-SOL as shown in Figure~\ref{fig:kstar_para_space}(a), 
(2) a fixed-fraction impurity model was employed to reduce computational costs by a factor of 20 compared to the multi-charged impurity model, 
and (3) the injected power is assumed to be evenly partitioned between electrons and ions.

This dataset extends over a significant part of KSTAR's typical operational space, specifically: $P_\text{inj}\in [1,8]$ MW, $n_\mathrm{e,core}\in [1.5,7]\times 10^{19}$ m$^{−3}$, $f_Z\in [0,0.04]$, $f_D\in [0.6,2.0]$, and $I_p \in [600,800]$ kA as illustrated in the shaded region of Figure~\ref{fig:kstar_para_space}(b). A sub-dataset of roughly 70,000 runs, named dataset A, is used for model training, as described in Section~\ref{sec:train}, while a separate dataset containing approximately 16,000 runs, named dataset B, is used for the evaluation of the combined model performance in Section~\ref{sec:perf}. A more detailed description of the UEDGE code setup and data generation can be found in Appendix~\ref{app:dataset}.

Apart from being used for surrogate model training and validation, this large dataset exhibits several interesting and robust statistical features, indicating intriguing physics underlying the divertor plasma that would not be available in previous case-by-case studies. \rev{For example, it reveals a bi-Gaussian distribution of the outer strike-point electron temperature $T_\mathrm{e,osp}$ (i.e., the local $T_\mathrm{e}$ on the outer divertor target just outside the separatrix)~\cite{zhao2024extended}, a feature that is qualitatively consistent with recent observations from ASDEX Upgrade~\cite{scotti2024existence}.} These findings will be reported and discussed in a separate paper.

\rev{It is worth noting that the fluid neutral model, which is only valid in the collisional limit, was used to generate our database. As a result, non-Maxwellian effects of neutrals \rrev{were not} captured by current database, and neutral particle transport, particularly in less collisional regions (e.g., higher temperature and lower density), may be overestimated compared to predictions from a kinetic neutral model. In detached divertor plasmas, the fluid neutral model is expected to perform well, as neutrals are typically collisional due to the low temperature and high density. However, the absence of a kinetic neutral description may introduce inaccuracies in predicting attached plasma behavior and the onset of detachment. This limitation will be addressed in our future study.}

\section{Training of surrogate model}\label{sec:train}

This section describes the two-stage training process of the DivControlNN model. The first stage involves the development and training of a multi-modal $\beta$-variational autoencoder ($\beta$-VAE) to identify a latent space representation (LSR), $\mathbf{z}$, for the 2D tokamak boundary plasma state. To enhance performance, the desired information, or the surrogate model outputs, $\mathbf{y}$, is categorized into three groups of plasma diagnostics - downstream plasma profiles, divertor radiation information, and upstream plasma profiles - using a specialized encoder architecture. The second stage focuses on training a multi-layer perceptron (MLP) to map the control parameters ($\mathbf{x}$) to the latent space representation ($\mathbf{z}$) identified by the $\beta$-VAE.

\subsection{Latent space identification}

\begin{figure}[h]
    \centering
    \includegraphics[width=0.8\linewidth]{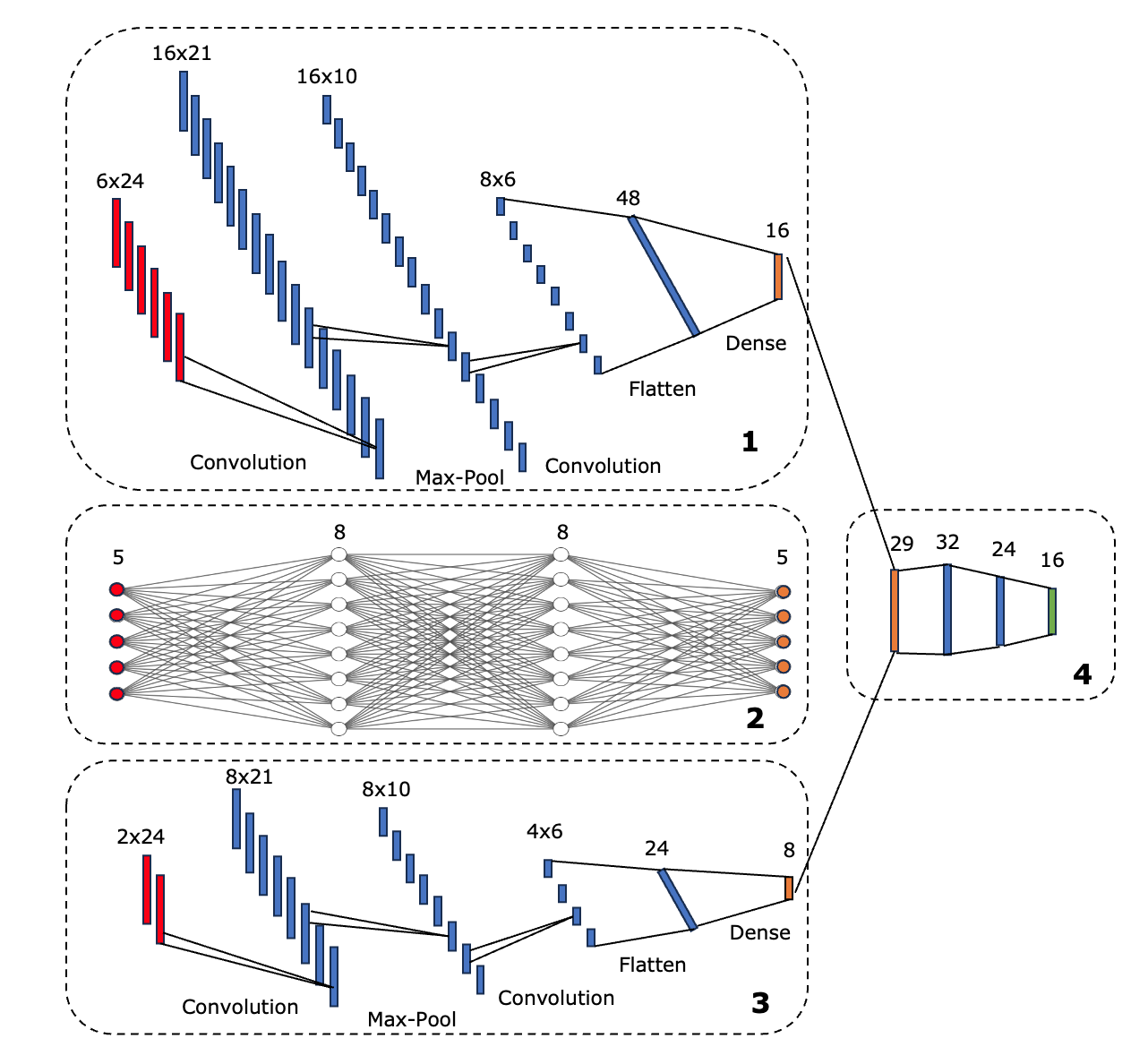}
    \caption{The architecture of the multi-modal encoder within the multi-modal $\beta$-VAE. It consists of two 1D CNNs (1 and 3) and one MLP (2), which process profiles and scalar information separately. The resulting intermediate latent space representations (LSRs, shown in orange) are then concatenated and further compressed using another MLP (4). The inputs are highlighted in red, while the final LSR is represented in green.}
    \label{fig:encoder}
\end{figure}

A multi-modal $\beta$ variational auto-encoder ($\beta$-VAE) is constructed and trained to identify an appropriate latent space representation (LSR) for the axisymmetric 2D tokamak boundary plasma state for detachment control. As mentioned in the previous section, the desired synthetic diagnostics or predictions can be categorized into three groups depending on their locations: (1) downstream plasma information on the divertor target plates (i.e., six one-dimensional arrays, each with 24 elements), (2) divertor radiation information (i.e., including \rev{overall and divertor} radiation fractions, peak radiation \rev{intensity and location in both radial and poloidal directions}, in total five scalars), and (3) upstream plasma information at the outboard midplane (i.e, two one-dimensional arrays, each with 24 elements). To enhance its performance, the multi-modal $\beta$-VAE processes these three groups of information separately.
\rev{The rationale behind this approach is that grouping the data based on their spatial locations and modalities minimizes interference between different regions and modalities, allowing the model to better capture localized patterns. Additionally, this approach offers the flexibility to adjust relative prediction accuracy by either modifying the dimensions of the corresponding latent space $\mathbf{z}_i$ or tuning the weights in the loss functions, thereby facilitating targeted optimization for specific regions or features.}
As shown in Figure~\ref{fig:encoder}, the encoder architecture consists of four sub-encoders: two 1D convolutional neural networks (CNNs) to pre-compress the one-dimensional profiles to latent space $\mathbf{z}_1$ and $\mathbf{z}_3$, one multi-layer perceptron (MLP) to project scalars to the latent space $\mathbf{z}_2$, and another MLP to further compress the combined latent variables $\{\mathbf{z}_i\}$. Specifically, after applying convolutions and max pooling in the CNNs, the dimensionalities of the pre-compressed profiles are set to 8 and 16 for the upstream and downstream plasma profiles, respectively, corresponding to pre-compression rates of 6 and 9. Together with the latent variable representing divertor region radiation information, these latent variables are concatenated and further compressed into a final 16-dimensional latent space representation (LSR) $\mathbf{z}$. This process, often referred to as latent space fusion, ensures that the final LSR $\mathbf{z}$ can self-consistently capture the essential features of multi-modal diagnostics. Latent space fusion is an active area of research, with new and more advanced techniques continually being developed~\cite{cannistraci2023bricks}, which will be explored in our future studies. Overall, this $\beta$-VAE achieves a moderate compression rate of approximately 12, as the degrees of freedom (DOFs) from the inputs total 197, while the final LSR dimension is 16.

Similarly, the decoder follows the same architecture as the encoder, consisting of four sub-decoders arranged in reverse order. Consequently, the total loss function comprises four components,
\begin{equation}\label{eq:loss}
    \mathcal{L}=\alpha_1\mathcal{L}\left(\mathbf{y'}_{1}-\mathbf{y}_{1}\right)+\alpha_2\mathcal{L}\left(\mathbf{y'}_{2}-\mathbf{y}_{2}\right)+\alpha_3\mathcal{L}\left(\mathbf{y'}_{3}-\mathbf{y}_{3}\right)+\beta\mathcal{D}_\mathrm{KL}.
\end{equation}
The first three terms in Equation~\ref{eq:loss} represent the difference between the reconstructed quantities, $\mathbf{y'}$, and the inputs, $\mathbf{y}$, for three groups of predictions, while the last term corresponds to the Kullback–Leibler (KL) divergence, which measures the likelihood of the distribution with respect to the reference distribution. Here, $\alpha_i$ and $\beta$ are user-defined loss weights. 

For the results presented in this paper, $\alpha_1=6,\alpha_2=\alpha_3=2$, and a small $\beta=10^{-7}$ is used to regularize latent space variable to a normal distribution. The mean squared error (MSE) is used as the loss functions $\mathcal{L}$, and the Adam optimizer is used for gradient descent. With the random sampling, the dataset is split into training, validation, and test sets with a ratio of 72\%:18\%:10\%. For dataset A, which contains 70,659 cases, this corresponds to 50,876 training cases, 12,718 validation cases, and 7,065 test cases. The batch size is set to 256, and the model is trained for 3,000 epochs. Both a fixed learning rate of 0.01 and an exponential decay learning rate schedule (e.g., an initial learning rate of 0.02 with a decay rate of 0.95 and decay steps of 10,000) were tested and found that they yield similar results (to be discussed in Section~\ref{subsec:uq}).

\begin{figure}[h]
    \centering
    \includegraphics[width=\linewidth]{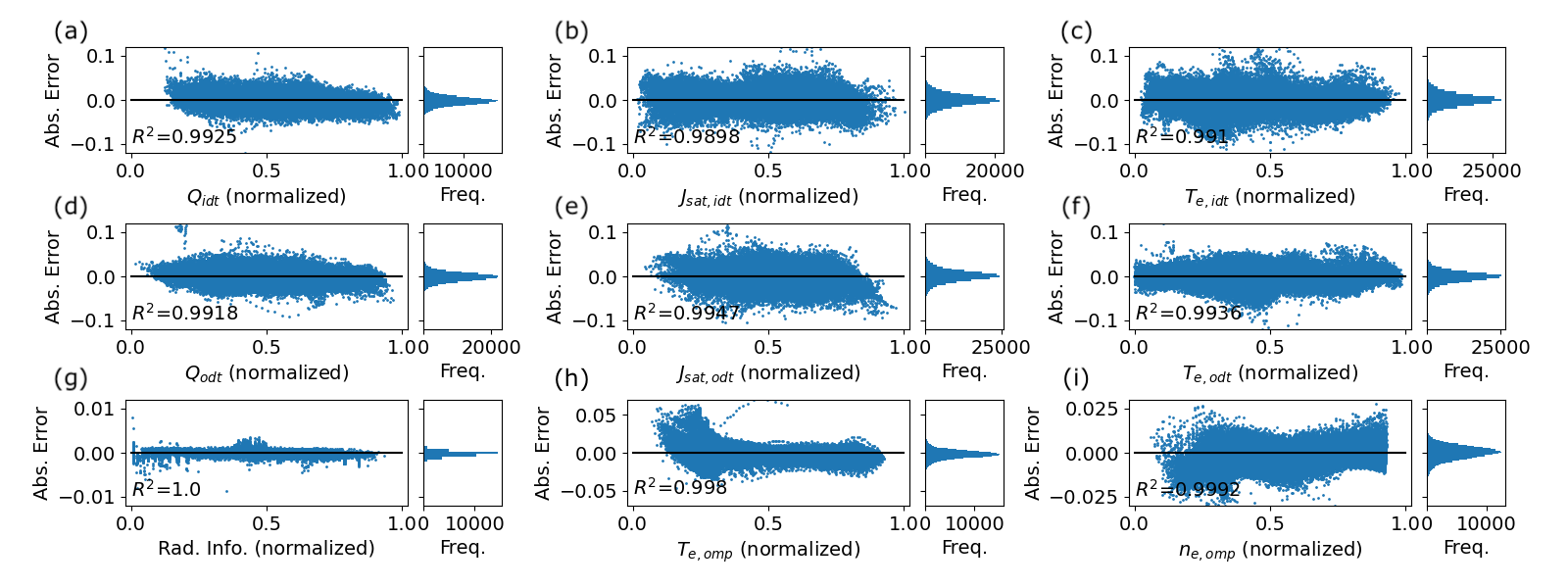}
    \caption{Performance of a typical multi-modal $\beta$-VAE evaluated in terms of absolute residual error versus the ground truth (in normalized units), along with histograms of the error distribution for all quantities in the test dataset\rev{, including (a) $Q_\mathrm{idt}$, (b) $J_\mathrm{sat,idt}$, (c) $T_\mathrm{e,idt}$, (d) $Q_\mathrm{odt}$, (e) $J_\mathrm{sat,odt}$, (f) $T_\mathrm{e,odt}$, (g) radiation information, (h) $T_\mathrm{e,omp}$, and (i) $n_\mathrm{e,omp}$.} }
    \label{fig:mmae_perf}
\end{figure}

As an example of the performance of multi-modal $\beta$-VAE, Figure~\ref{fig:mmae_perf} shows the \rev{absolute} error between $\beta$-VAE reconstructed outputs and inputs versus the ground truth (\rev{i.e., quantities such as 1D heat load, ion saturation current, and electron temperature profiles on the inner and outer divertor targets, divertor radiation information, as well as 1D electron density and temperature profiles at the outboard midplane,} all expressed in normalized units) for the test dataset (i.e., 7,065 cases) of one training trial. In this Figure, all 24 elements in the one-dimensional array are plotted together, resulting in approximately 170,000 data points per plot, except for Figure~\ref{fig:mmae_perf}(f), which contains roughly 35,000 data points. While the scatter plots may appear to show a wide range of errors due to the large number of data points, the histograms of the error distribution reveal that the spread of errors is relatively narrow, with the majority of cases exhibiting very small errors. Additionally, the $R^2$ score (coefficient of determination) for all quantities exceeds 0.98, indicating near-perfect predictive performance from a statistical perspective.

\begin{figure}[h]
    \centering
    \includegraphics[width=\linewidth]{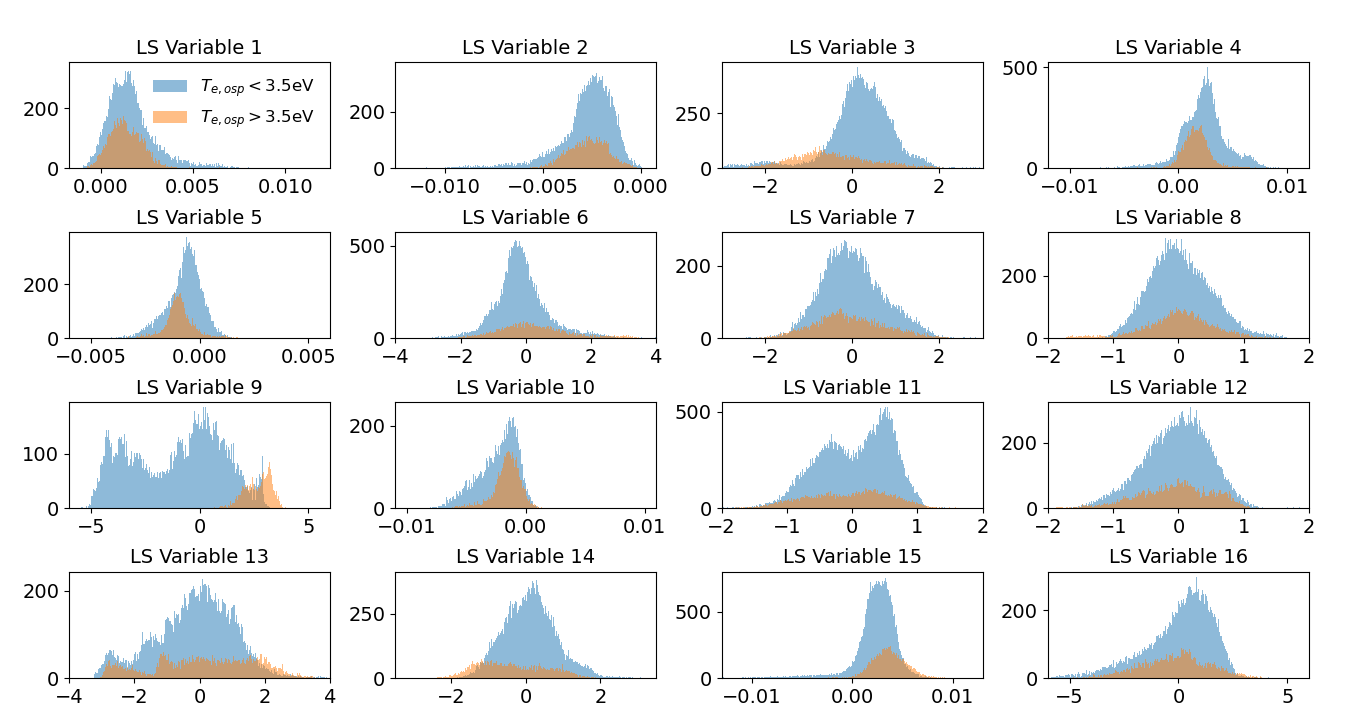}
    \caption{Histograms of each dimension of the LSR for dataset A. The $x$-axis represents the latent space values, while the $y$-axis denotes the number of cases or frequency.}
    \label{fig:lsr_dist}
\end{figure}

The resulting 16-dimensional LSR for all cases in dataset A from one selected training trial is shown in Figure~\ref{fig:lsr_dist}. Here, all cases are labeled as either ``detached" or ``attached" based on whether the electron temperature at the outer divertor strike-point is below or above 3.5 eV. Under this classification, 51,459 out of 70,659, or approximately 72.83\% of cases are detached in the training dataset A. It is important to note that this is a simplified classification of divertor detachment, and the label is used solely to highlight potential discrepancies in the latent space representations for different scenarios. This labeling does not influence the model training process and, therefore, has no impact on the model's performance. 
From Figure~\ref{fig:lsr_dist}, most of the latent variables for detached and attached cases span the same value range, and their distributions are qualitatively similar. However, an exception is observed in one latent variable (i.e., variable 9), where attached cases tend to exhibit positive values, while detached cases tend to have smaller or even negative values. This suggests that this latent space variable might be the most impactful latent dimension for distinguishing outer divertor detachment based on the strike-point electron temperature criterion.

\begin{figure}[h]
    \centering
    \includegraphics[width=0.49\linewidth]{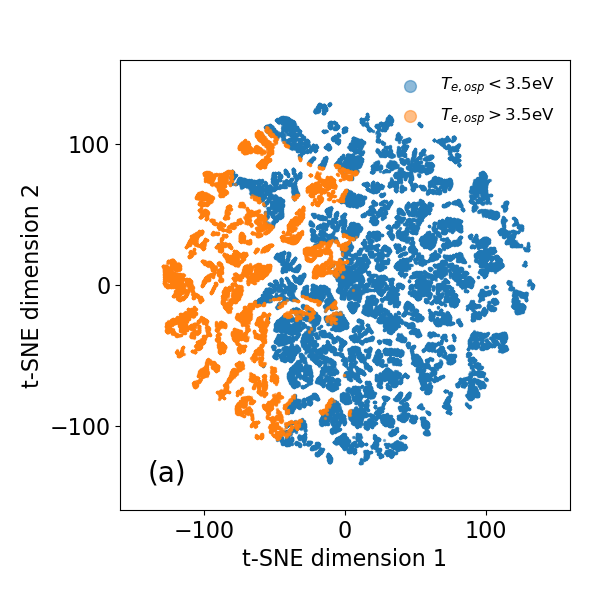}
    \includegraphics[width=0.49\linewidth]{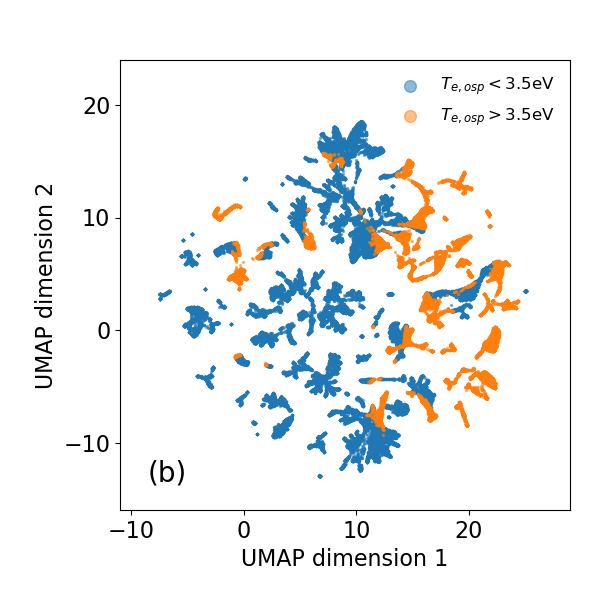}
    \caption{Visualization of the resulting LSR for the 2D boundary plasma system using (a) t-SNE \rev{with perplexity coefficient of 50}, and (b) UMAP \rev{with parameters set to \texttt{n\_neighbors}=15, \texttt{n\_components}=2, and \texttt{random\_state}=10}.}
    \label{fig:tsne_umap}
\end{figure}

\rev{As illustrated in Figure~\ref{fig:tsne_umap},} t-Distributed Stochastic Neighbor Embedding (t-SNE)~\cite{van2008visualizing} and Uniform Manifold Approximation and Projection (UMAP)~\cite{mcinnes2018umap} are utilized to better understand and visualize the 16-dimensional latent space representation. Both methods reduce the high-dimensional data to 2D while preserving the local relationships within the data. Unlike the 1D surrogate models, where detached and attached cases are well-separated~\cite{zhu2022data,holt2024tokamak}, in this 2D representation, the boundary between detached and attached cases is less distinct, although they still generally occupy different regions. 
This difference arises because, in a 1D flux tube, divertor plasmas are limited to two possible scenarios, either detached or attached, aligning with the discrete, binary labels. However, in a 2D system, the additional dimension along the radial direction significantly expands the range of possible scenarios. For instance, instead of being strictly in a detached or attached state (for all radial locations), the plasma can also be in a partially detached state. Furthermore, the inner and outer divertor regions may be in different states, adding further complexity to the classification. In other words, although the training data are labeled with binary classifications, the actual features of these data are non-binary and likely continuous. The t-SNE and UMAP plots, therefore, highlight the necessity of transitioning from a 1D to a 2D surrogate model for practical applications.

\subsection{Latent space mapping}

With the proper latent space identified and the corresponding LSR $\mathbf{z}$ obtained for the entire training dataset A, it becomes straightforward to train a multi-layer perceptron (MLP) that maps the control parameters $\mathbf{x}$ to $\mathbf{z}$.
In this study, the input $\mathbf{x}$ consists of five key discharge parameters, as explained in Section~\ref{sec:prep}, and the output latent space representation (LSR) $\mathbf{z}$ is 16-dimensional. An MLP with four hidden layers, 32 neurons per layer, was constructed. The neural network utilized the ReLU activation function except for the output layer, the Adam optimizer, and the mean squared error (MSE) loss function. An exponential decay schedule was applied for the learning rate, starting with an initial value of 0.01, a decay rate of 0.95, and a decay step of 50,000. Consistent with the multi-modal $\beta$-VAE training, dataset A was randomly divided into training (72\%), validation (18\%), and test (10\%) sets. The MLP was trained for 2,000 epochs, with the best model (i.e., the minimum validation loss) selected during training.

\begin{figure}[h]
    \centering
    \includegraphics[width=\linewidth]{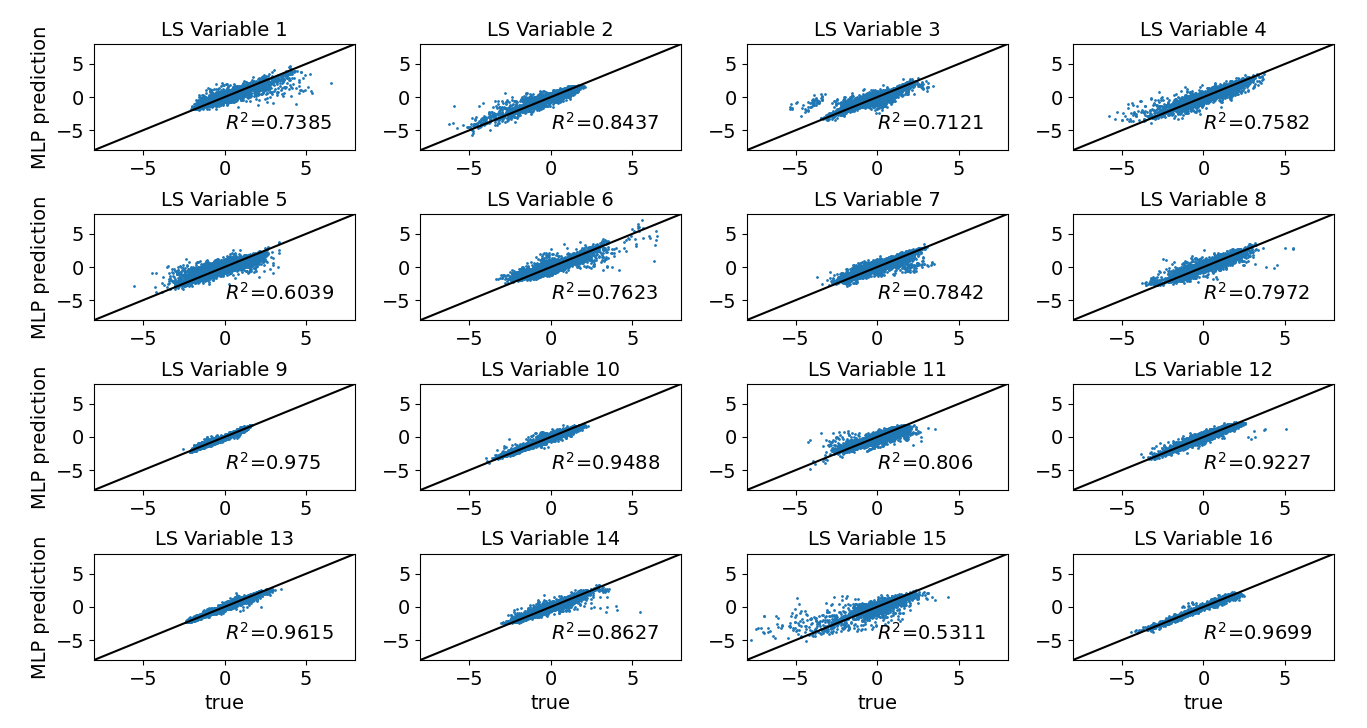}
    \caption{Performance of an MLP model evaluated in terms of predictions ($y$-axis) versus true values ($x$-axis) for all sixteen regularized latent variables, based on the LSR $\mathbf{z}$ obtained in Figure~\ref{fig:mmae_perf}. The black lines ($y = x$) represent the ideal predictions with no error.}
    \label{fig:mlp_perf}
\end{figure}

Noticing that some latent variables may be one or two orders of magnitude smaller than the rest (as shown in Figure~\ref{fig:lsr_dist}), we standardized the latent variable distributions to ensure a more uniform prediction across all latent variables by the MLP. The standardization was performed using the formula $f_S=\left(f-\bar{f}\right)/\sigma_f$, where $\bar{f}$ and $\sigma_f$ are the mean and standard deviation of the distribution $f$. 
Nevertheless, the accuracy of the MLP predictions, even after standardization, as shown in Figure~\ref{fig:mlp_perf}, often correlates with the original amplitude of the latent space range. Latent dimensions with smaller value ranges, such as variables 5 and 15, tend to exhibit worse accuracy, whereas dimensions with larger value ranges, such as variables 9, 13, and 16, generally achieve better accuracy.
Interestingly, the standardization process appears to have minimal impact on the performance of the combined surrogate model (see \rev{Tables~\ref{tab:model_perf_stat_div}-\ref{tab:model_perf_stat_omp_regularized}} and \rev{Figure~\ref{fig:combined_perf}} in Appendix~\ref{app:perf}). This suggests that these small-valued latent space dimensions are likely more strongly influenced by the regularization process during $\beta$-VAE training and may play less significant roles in characterizing the overall boundary plasma states.

\section{Performance of surrogate model}\label{sec:perf}

In this section, we first evaluate the accuracy of the surrogate model, DivControlNN, against an independent dataset and quantify its uncertainty using the deep ensemble method. Next, we highlight the significant computational speed-up (over $10^8$) achieved by the model, which is critical for real-time applications such as detachment control. Finally, the model's practical performance is demonstrated through several examples, including detachment onset prediction and broader divertor plasma dynamics, showcasing its accuracy, robustness, and efficiency in comparison to UEDGE simulations.

\subsection{Accuracy}\label{subsec:accuracy}

Following the two-stage training process for latent space identification and mapping, as described in the previous section, the resulting surrogate model, DivControlNN, which integrates the MLP and the multi-modal decoder, is ready for evaluation. Its accuracy is evaluated using a separate dataset (i.e., dataset B), consisting of 16,293 converged UEDGE simulations that are independent of the training dataset (i.e., dataset A) used during surrogate model training. Although dataset B is independent, it spans the same parameter space as dataset A and exhibits similar distributions in terms of sampling rate and overall plasma solutions (e.g., the ratio of detached to attached cases). This similarity minimizes biases in performance evaluation caused by potential distribution shifts.

\begin{figure}[h]
    \centering
    \includegraphics[width=0.6\linewidth]{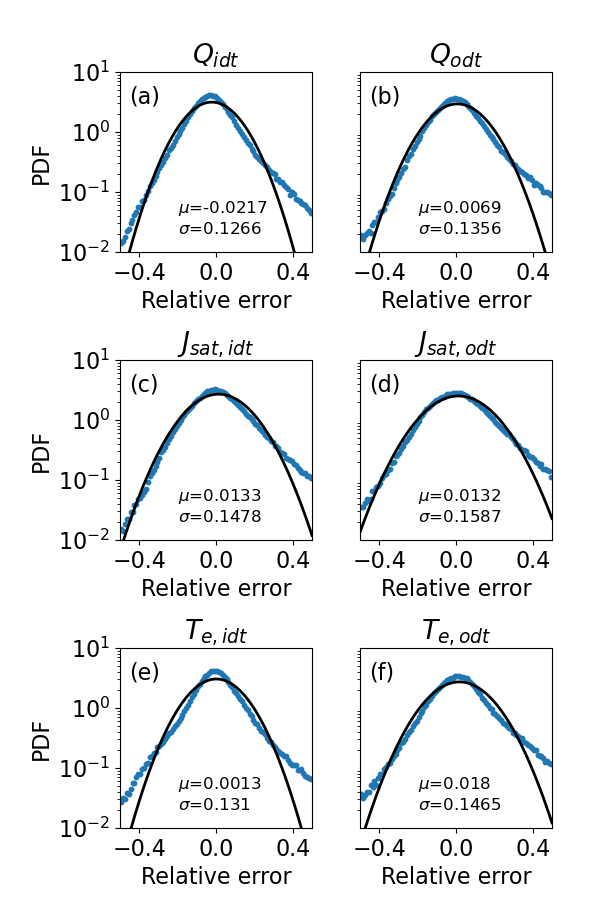}
    \caption{Probability Density Functions (PDFs) of surrogate model prediction errors for plasma profiles at the inner \rev{(left panel)} and outer \rev{(right panel)} divertor targets (denoted by the subscripts \textit{idt} and \textit{odt}, respectively), including the heat flux $Q$ \rev{(a, b)}, ion \rev{saturation} current $J_{sat}$ \rev{(c, d)}, and electron temperature $T_e$ \rev{(e, f)}. Black lines represent the corresponding normal distribution fitted with the mean $\mu$ and standard deviation $\sigma$.}
    \label{fig:pdf_div_model9}
\end{figure}

\begin{figure}[h]
    \centering
    \includegraphics[width=0.6\linewidth]{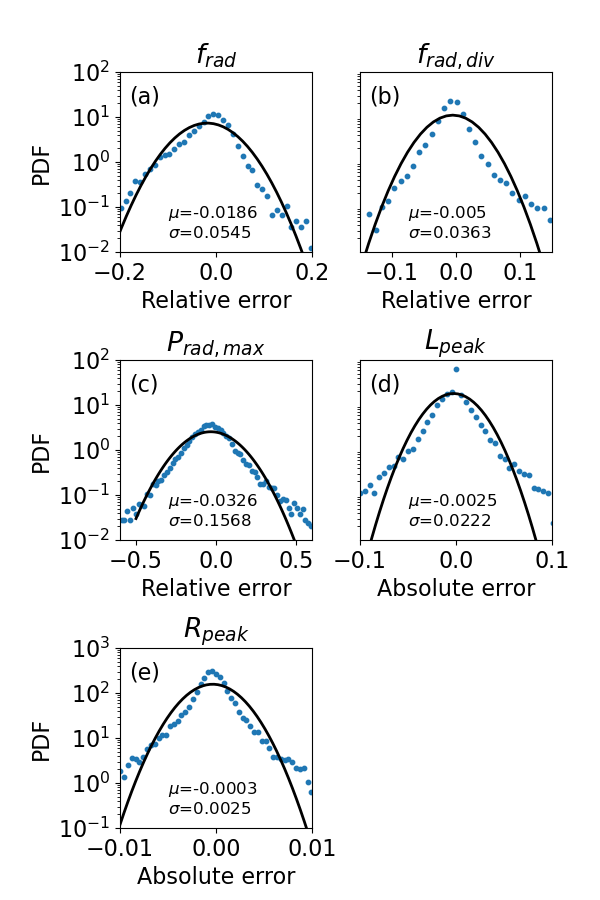}
    \caption{Probability Density Functions (PDFs) of surrogate model prediction errors for radiation information at divertor region, including \rev{(a)} the edge radiation fraction $f_\mathrm{rad}$, \rev{(b)} divertor radiation fraction $f_\mathrm{rad,div}$, \rev{(c)} peak radiation power $P_\mathrm{rad,max}$, and the location of peak radiation at out divertor region (i.e., \rev{(d)} poloidal location $L_\mathrm{peak}$ and \rev{(e)} radial location $R_\mathrm{peak}$). The black lines represent the corresponding normal distribution fitted with the mean $\mu$ and standard deviation $\sigma$. }
    \label{fig:pdf_rad_model9}
\end{figure}

\begin{figure}[h]
    \centering
    \includegraphics[width=0.6\linewidth]{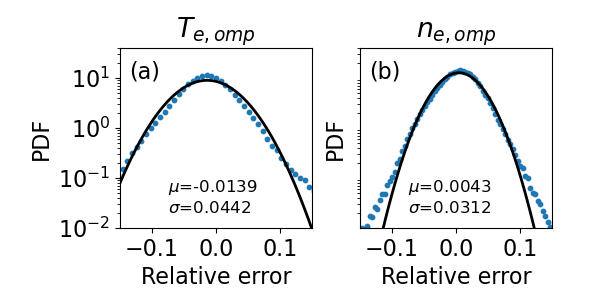}
    \caption{Probability Density Functions (PDFs) of surrogate model prediction errors for \rev{(a)} upstream plasma temperature $T_\mathrm{e,omp}$ and \rev{(b)} density $n_\mathrm{e,omp}$ profiles at the outboard midplane, with the black lines representing the corresponding normal distribution fitted with the mean $\mu$ and standard deviation $\sigma$.}
    \label{fig:pdf_omp_model9}
\end{figure}

The accuracy of the surrogate model is illustrated in Figures~\ref{fig:pdf_div_model9} to~\ref{fig:pdf_omp_model9}, where predictions of key plasma properties, all expressed in physical units, are evaluated based on their error distributions. In these figures, the corresponding normal distributions with the mean ($\mu$) and standard deviation ($\sigma$) based on the respective error distributions are also plotted for reference. While most error distributions do not perfectly align with the normal distributions, as they often exhibit more pronounced central peaks and fatter tails (i.e., resembling a super-Gaussian distribution), they remain close enough to the normal distributions to allow $\sigma$ to be used as a good proxy for indicating the confidence interval. For all predicted quantities (except for the peak radiation position prediction, which is evaluated in absolute error), the mean relative error ($\mu$) is within a few percent or less, and the standard deviation ($\sigma$) varies between 3\% and 16\%. Notably, $\sigma$ is always at least an order of magnitude larger than $\mu$, indicating that the error distributions are dominated by variability rather than systematic bias. This suggests that the surrogate model provides accurate predictions with a well-defined range of uncertainty for relevant plasma properties.

\begin{figure}[h]
    \centering
    \includegraphics[width=0.6\linewidth]{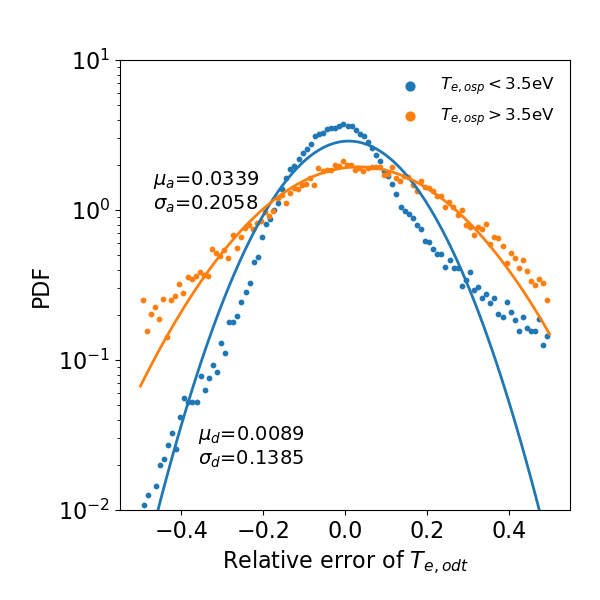}
    \caption{Probability Density Functions (PDFs) of surrogate model prediction errors for outer divertor electron temperature $T_\mathrm{e,odt}$ with $\psi_n\in[1.0,1.01]$ for detached (i.e., $T_\mathrm{e,osp}<3.5$ eV, in blue) and attached (i.e., $T_\mathrm{e,osp}>3.5$ eV, in orange) cases. The solid lines denote the corresponding normal distributions fitted with the mean $\mu$ and standard deviation $\sigma$. }
    \label{fig:pdf_teodt_model9}
\end{figure}

More specifically, for the downstream profile prediction (Figure~\ref{fig:pdf_div_model9}), the accuracy is quite similar between the inner and outer divertor target plates for all three quantities (i.e., heat load, ion saturation current, and electron temperature), with $\sigma$ ranging from approximately 12\% to 16\%. A more detailed analysis reveals that for quantities of particular interest, such as the electron temperature at the outer divertor plate in the near scrape-off layer (i.e., $\psi_n \in [1.0, 1.01]$, as illustrated in Figure~\ref{fig:pdf_teodt_model9}), the error distribution, and hence the model accuracy, can vary between detached and attached states (i.e., labeled with the outer strike point temperature below or above 3.5 eV), with $\sigma_\mathrm{detached}<\sigma_\mathrm{combined}<\sigma_\mathrm{attached}$. This is likely due to the larger number of detached cases in our training and validation datasets. However, the accuracy in both states remains close to the combined $\sigma$, i.e., without any order-of-magnitude discrepancy. 

For the radiation information in the divertor region (Figure~\ref{fig:pdf_rad_model9}), including the radiation fraction, divertor radiation fraction, peak radiation intensity, and location, the predicted results have varying accuracy. The radiation fraction and divertor radiation fraction have relatively small standard deviations (i.e., 5.45\% and 3.63\%, respectively), whereas the peak radiation intensity shows the largest standard deviation (e.g., 15.68\% in this case). This is likely due to the localized and highly nonlinear nature of the peak radiation intensity, which is more sensitive to the details of the steady-state divertor plasma solution. The position of peak radiation location on the outer divertor region, quantified by $L_\mathrm{peak}$ and $R_\mathrm{peak}$, is expressed in physical unit (i.e., meter). Here $L_\mathrm{peak}$ represents the poloidal distance measured from the outer divertor target along the corresponding magnetic flux surface so that $L_\mathrm{peak}$ is always positive. In contrast, $R_\mathrm{peak}$ represents the radial distance relative to the separatrix location, where a positive $R_\mathrm{peak}$ value indicates a position in the scrape-off layer (SOL), and a negative $R_\mathrm{peak}$ value corresponds to a location within the private flux region or the closed flux surface region. The relatively small standard deviations for $L_\mathrm{peak}$ and $R_\mathrm{peak}$ (i.e., $\sigma = 0.0222$ in the poloidal direction and $\sigma = 0.0025$ in the radial direction) indicate that the position prediction is quite accurate. 

Moreover, the upstream profile prediction (Figure~\ref{fig:pdf_omp_model9}) demonstrates greater accuracy than the downstream profiles, with $\sigma$ less than 5\% for the electron density and temperature profiles at the outboard midplane. These results demonstrate that while the surrogate model performs well across a range of plasma properties, the accuracy varies depending on the spatial region and the specific quantity being predicted.

Many factors influence the accuracy of a surrogate model. With fixed inputs $\mathbf{x}$ and outputs $\mathbf{y}$, the number of trainable parameters or the model's architecture plays a significant role. Generally, more complex architectures with higher degrees of freedom lead to greater accuracy. However, this improved accuracy often comes at the expense of reduced computational speed, which may impose constraints on practical applications. Furthermore, a more complex model necessitates a larger and more diverse training dataset, as the quality and variety of the data directly affect the model's ability to generalize effectively. In our study, the surrogate model leverages latent space (LS) to ensure consistent predictions across multiple observations, and the dimensionality of the LS $n_z$ dictates how much information can be retained. A larger LS dimensionality typically results in higher accuracy, while a smaller dimensionality can lead to a loss of critical information. For instance, the multi-modal $\beta$-VAE with 12 LS dimensions achieves sub-optimal performance compared to the version with 18 LS dimensions, highlighting the importance of selecting an appropriate LS size. Furthermore, the loss weight ratio used during the training of the multi-modal $\beta$-VAE significantly affects the accuracy of predictions for different quantities, especially when balancing contributions from multiple outputs.

Beyond model architecture and training parameters, the complexity of the underlying physics also plays a critical role in prediction accuracy. For example, outboard midplane profiles are generally predicted with higher accuracy, likely because these quantities are more closely tied to the global discharge parameters (i.e., model inputs). In contrast, downstream profiles are more challenging to predict due to the highly nonlinear dissipation, such as particle, momentum, and energy losses associated with atomic processes, occurring along magnetic field lines. This inherent complexity may also explain why the prediction of peak radiation intensity exhibits the largest relative error among all quantities. 

\subsection{Uncertainty quantification}\label{subsec:uq}


Although the accuracy of the surrogate model can be evaluated using an independent test dataset and quantified through statistical analysis, real-world applications present a different challenge. In scenarios where ground truth data is unavailable (e.g., detachment control), model uncertainty must be estimated through alternative means. More critically, the model’s predictive uncertainty is region and input dependent due to nonlinear plasma dynamics. For example, predictions for upstream conditions (e.g., midplane density, temperature) are generally more reliable, whereas downstream quantities (e.g., divertor electron temperature, peak radiation intensity) display broader spreads due to stronger dissipation effects and sensitivity to impurity radiation. Moreover, the model performance varies across input parameter space. 
Here, the term ``uncertainty" refers specifically to the inherent uncertainty arising from the surrogate model itself, not the uncertainty or measurement error associated with the model inputs. In other words, once trained, the surrogate model operates deterministically, where each set of inputs corresponds to a single, fixed set of outputs. However, it is still possible to estimate the model's uncertainty using the deep ensemble technique~\cite{lakshminarayanan2017simple}. The key idea behind deep ensembles is to construct an ensemble of diverse models that collectively represent a distribution over all possible models. Consequently, the ensemble of predictions from these models can be interpreted probabilistically, providing a measure of uncertainty and a level of confidence in the predictions.

Following this approach, seventy-two training trials were conducted for the multi-modal $\beta$-VAE training, each initialized with different random seeds. The random initialization of weights and biases defines different \rev{starting} point in parameter space, further influencing the optimization trajectory. This can potentially \rev{lead} to different convergence outcomes, where models are most likely \rev{settling} into distinct minima in the optimization landscape.
With all the other settings remaining the same, half of the trials (i.e., thirty-six) used a fixed learning rate of 0.01, while the other half employed an exponential decay learning rate. After excluding trials that exhibited overfitting or underfitting, the top ten best-performing trials were selected from the fifty-seven valid trials for the second stage of latent space mapping. In this stage, twenty trials for forward MLP training - half mapping model inputs to the native LSRs and the other half mapping to the regularized LSRs - were carried out, again with random initializations, for each selected multi-modal $\beta$-VAE (or identified latent space). The best-performing MLPs (one with native LSR and one with regularized LSR for each multi-modal $\beta$-VAE) were selected. In the end, two ensembles of surrogate models were attained as the combined models, one utilizing the native LSRs and the other using the regularized LSRs, with each ensemble consisting of ten independent surrogate models. 
Despite achieving overall similar accuracy, the individual models in the ensemble, trained with different seeds, often exhibit variations in their loss distributions, reflecting the structured minima in the optimization landscape.
The combined model, therefore, can provide an ensemble of interdependent predictions, where the variation in predictions can serve as a proxy for surrogate model uncertainty.

\begin{figure}[h]
    \centering
    \includegraphics[width=0.49\linewidth]{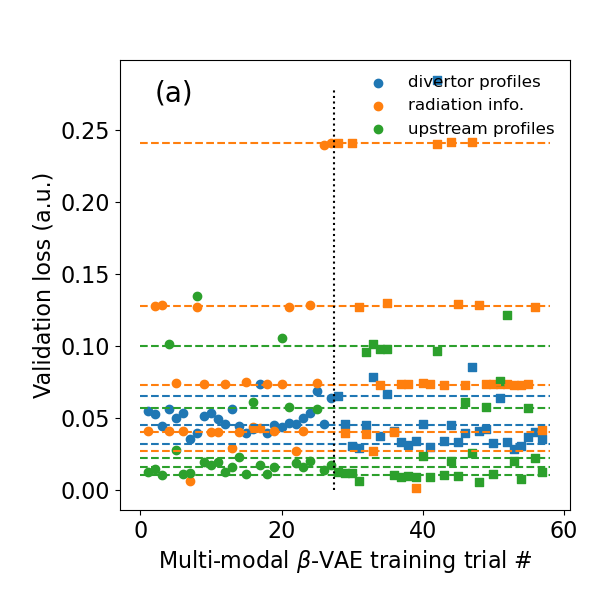}
    \includegraphics[width=0.49\linewidth]{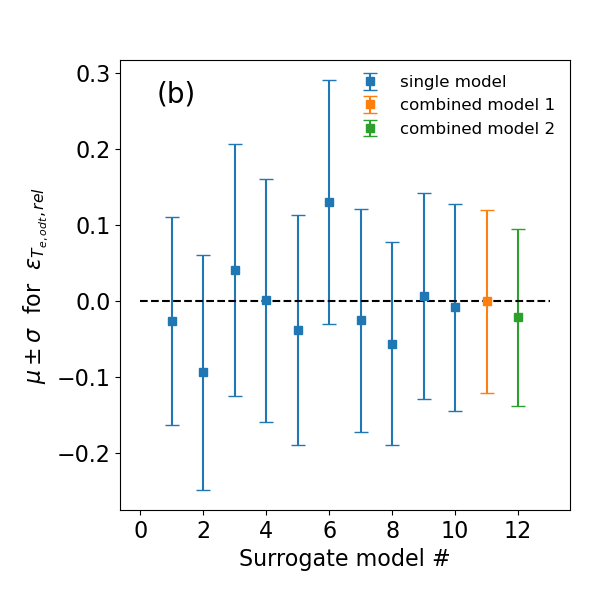}
    \caption{(a) Validation loss of each component from the multi-modal $\beta$-VAE across a total 57 trials. Trials 1-27 (in circles) were trained with a fixed learning rate of 0.01, while trials 28-57 (in squares) were trained using an exponential decay learning rate. (b) The performance of the surrogate model for the ten best surrogate models, along with the combined models, evaluated in terms of the mean and standard deviation of the relative error for predicting the outer divertor electron temperature profile using an independent test data set.}
    \label{fig:multi_loss}
\end{figure}

The trials with a fixed learning rate and those with an exponential decay learning rate yielded overall similar accuracies, as shown in Figure~\ref{fig:multi_loss}(a). Interestingly, it was observed that the training and validation losses of valid multi-modal $\beta$-VAE models were not uniformly distributed across all trials. Instead, they appeared to be discretized around specific values. These discretized values, indicated by the dashed lines in the plot, likely correspond to different minima in the parameter space of the machine learning models. If only a single model were to be selected, the one with the smallest losses (e.g., trial \# 7 or trial \# 39) would likely be preferred, as it is closer to the global minimum or optimal solution. However, for a deep ensemble, trials with sub-optimal performance are also acceptable. 

Besides providing uncertainty quantification, the combined model, i.e., taking the mean of all predictions across the models within the ensemble, often achieves better accuracy. In our study, the combined models demonstrate improved accuracy, as reflected by smaller standard deviations of prediction errors for most quantities, except for certain divertor radiation information (e.g., the peak radiation intensity) when validated against dataset B. For example, as shown in Figure\ref{fig:multi_loss}(b), the standard deviation of the relative error in electron temperature predictions at the outer divertor decreases from 15–18\% in individual models to approximately 13\% in combined models. Interestingly, an alternative combined model consisting of only the top five best-performing models (referred to as the combined model 2) shows similar performance to the one using all ten surrogate models (referred to as the combined model 1). This suggests that the deep ensemble technique does not heavily rely on the individual performance of each model. Detailed performance metrics for individual and combined models are provided in Appendix~\ref{app:perf}.

\subsection{Speed}

The surrogate model, DivControlNN, is specifically designed for real-time control of divertor detachment, making speed a critical metric for its real-time capabilities. Like most machine learning models, DivControlNN is implemented in Python and consists of two neural networks (i.e., a multi-model decoder and a forward prediction multi-layer perceptron). These neural networks are built and trained using the \texttt{TensorFlow} platform~\cite{tensorflow2015-whitepaper} on the GPU node of NERSC's Perlmutter system. Each GPU computational node features one AMD EPYC 7763 CPU and four NVIDIA A100 GPUs. \rev{The typical training time for the $\beta$-VAE is approximately 45 hours, while the MLP requires only about 15 minutes.}

After training, the Python implementation of DivControlNN, without any optimization, achieves a prediction time of approximately 90 ms on NERSC's computational node when using only the CPU. For comparison, a typical UEDGE simulation with this setup could take several hours to reach a steady-state solution, assuming it converges. Therefore, even in its unoptimized Python form, DivControlNN is approximately 100,000 times faster than direct numerical simulations. However, this speed is still insufficient for real-time control applications, where prediction times must be reduced to microsecond levels.

To address this limitation, the surrogate model is converted to C using the \texttt{keras2c} library~\cite{conlin2021keras2c}. After conversion, DivControlNN achieves significantly faster prediction times of 193 $\mu$s and 235 $\mu$s per inference on General Atomics' Iris system (equipped with Intel Xeon E5-2630 v3 2.40GHz CPUs) and KSTAR's ksim2 system (i.e., KSTAR Plasma Control System's testing computer which is equipped with Intel Xeon E5-2630 v2 2.60GHz CPUs), respectively. These improvements bring the model to real-time control requirements, achieving an overall speed-up on the order of $O(10^8)$ compared to the traditional UEDGE simulations. Furthermore, prediction times can be further reduced with more advanced CPUs. For instance, testing on Apple's M2 Pro processor demonstrates a prediction time of 87 $\mu$s, or just 18 $\mu$s when the \texttt{-Ofast} compiler optimization flag is enabled, highlighting the potential for even greater performance with modern hardware.

\subsection{Examples}

To further demonstrate DivControlNN’s capability, we present several examples of direct comparisons between surrogate model predictions and UEDGE simulation results. We first focus on detachment onset prediction, as this is a critical operational point where the controller is most needed.

\begin{figure}[h]
    \centering
    \includegraphics[width=\linewidth]{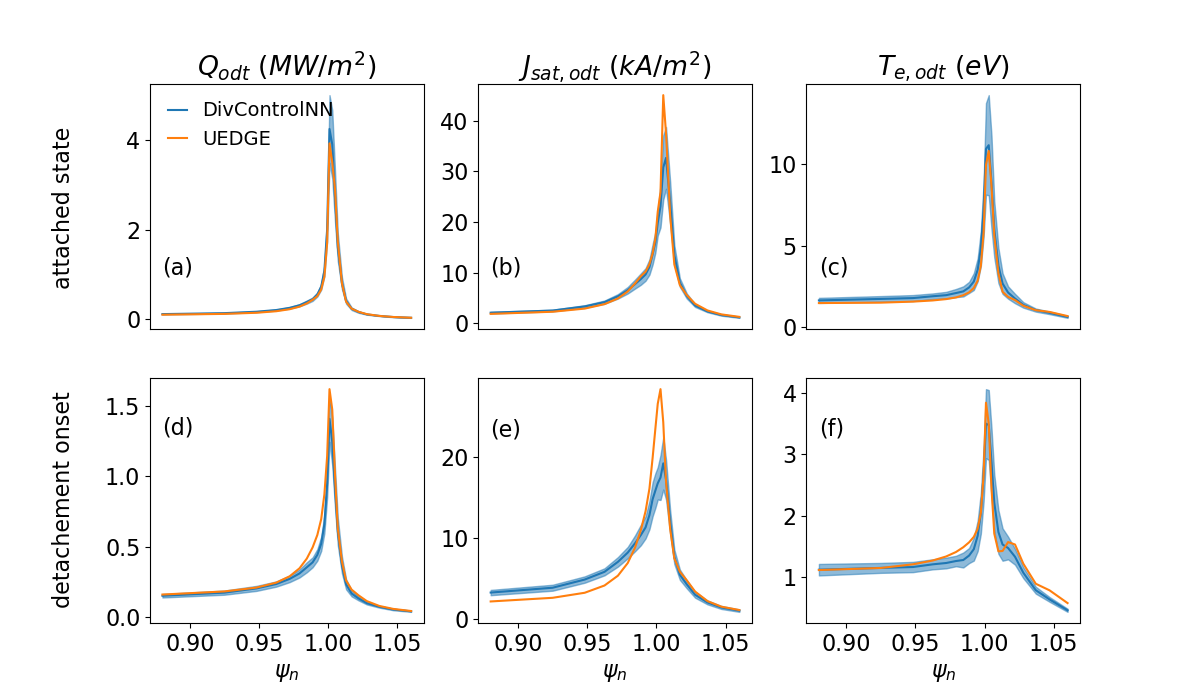}
    \caption{Examples of DivControlNN predictions (blue lines with shaded regions representing uncertainty) compared to UEDGE simulation results (orange lines) for outer divertor plasma conditions before (top panel) and near (bottom panel) the onset of detachment. \rev{The compared quantities include profiles of heat load (a, d), ion saturation current density (b, e), and electron temperature (c, f).} Both cases share the same discharge conditions (i.e., $I_\mathrm{p}$=700 kA, $P_\mathrm{inj}$=5.9 MW, $n_\mathrm{e,core}=3.55 \times 10^{19}\mathrm{m}^{-3}$, and $f_D=1.3$) except the impurity concentration: the attached state has $f_Z=$0.57\% while the onset case has $f_Z=$1.12\%. }
    \label{fig:onset}
\end{figure}

As illustrated in Figure~\ref{fig:onset}, for the selected discharge parameters (e.g., plasma current at 700 kA, total injection power at 5.9 MW, core plasma density at $3.55 \times 10^{19}\mathrm{m}^{-3}$, and a diffusivity scaling factor $f_D=1.3$), increasing the impurity fraction from 0.57\% to 1.12\% leads to significant changes of downstream plasma state at the outer divertor target plate. Specifically, the peak heat load drops from approximately 4 MW/m$^2$ to around 1.5 MW/m$^2$, the ion saturation current falls from about 40 kA/m$^2$ to roughly 25 kA/m$^2$, and the outer strike point electron temperature decreases from just above 10 eV to approximately 3.5 eV, indicating the onset of detachment. In Figure~\ref{fig:onset}, the blue and orange lines represent the DivControlNN model prediction and UEDGE simulation results, respectively, while the shaded blue area shows the uncertainty estimate from the deep ensemble method (i.e., the standard deviation from the prediction ensemble). The surrogate model successfully predicts detachment onset, as its predictions align closely with the UEDGE results. It captures the overall profiles of all three quantities and predicts the peak locations closely to the UEDGE simulation results despite minor deviations in absolute values.

\begin{figure}[h]
    \centering
    \includegraphics[width=0.8\linewidth]{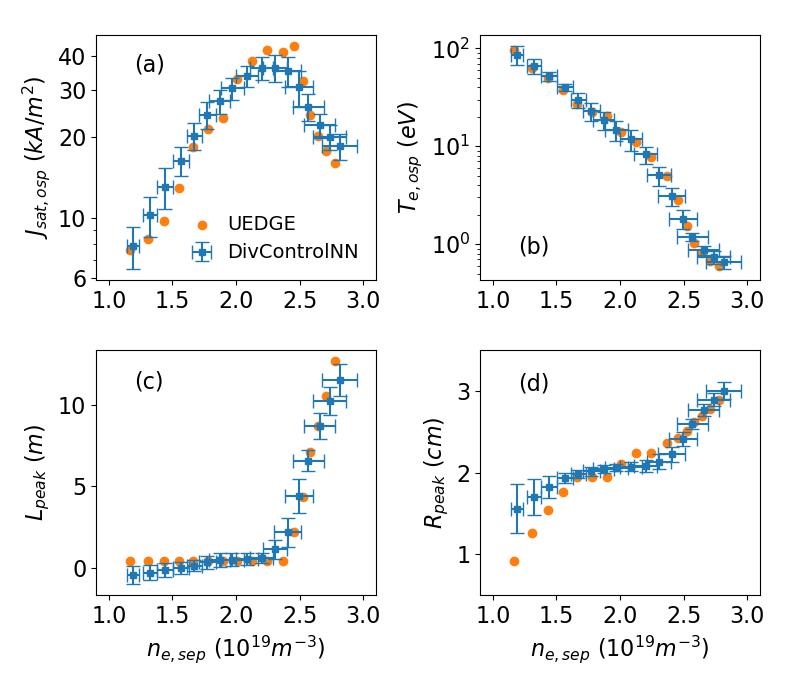}
    \caption{Examples of DivControlNN predictions and the corresponding UEDGE simulations results of (a) ion saturation current density, (b) electron temperature at the outer strike point, (c) poloidal and (d) radial locations of peak radiation in the outer divertor region.}
    \label{fig:rollover}
\end{figure}

In addition to specific case predictions near the detachment onset, the surrogate model successfully reproduces broader divertor plasma dynamics, including the overall trends observed in the UEDGE model. For example, Figure~\ref{fig:rollover}(a) demonstrates that DivControlNN accurately predicts the rollover behavior of the outer divertor ion saturation current at the strike point, $J_\mathrm{sat,osp}$, as a function of the electron density at the separatrix, $n_\mathrm{e,sep}$, i.e., the common density-scan exercise used to determine detachment. 
In this scan, the plasma current is $I_\mathrm{p} = 800$ kA, the total injected power is $P_\mathrm{inj} = 4.97$ MW, the impurity fraction is $f_Z = 0.56\%$, the diffusivity scaling factor is $f_D = 0.87$, and the core plasma density increases from $2.44$ to $6.90 \times 10^{19}\mathrm{m}^{-3}$. When the core plasma density reaches $n_\mathrm{e,core} = 5.23 \times 10^{19}\mathrm{m}^{-3}$, corresponding to a separatrix density of $n_\mathrm{e,sep} = 2.37 \times 10^{19}\mathrm{m}^{-3}$, the ion saturation current begins to decrease as the separatrix (or core) plasma density increases (i.e., rollover), indicating the onset of detachment. \rev{Consequently, the electron temperature at the outer strike point, $T_\mathrm{e,osp}$, as shown in Figure~\ref{fig:rollover}(b) begins to drop below 5 eV.} While few individual DivControlNN $J_\mathrm{sat,osp}$ prediction points appear to deviate slightly from the UEDGE simulation data, the overall trend is well recovered.
Furthermore, DivControlNN effectively captures the evolution of the poloidal and radial locations of the peak radiation as shown in Figure~\ref{fig:rollover}(c) and (c). When the plasma is in the attached state (e.g., $n_\mathrm{e,sep}<2.37\times10^{19}\mathrm{m}^{-3}$), the peak radiation is located close to the separatrix radially (i.e., $R_\mathrm{peak} \lesssim 2$ cm) and near the target plate poloidally (i.e., $L_\mathrm{peak}\simeq 0$). Once detachment occurs, the peak radiation location begins to shift radially further outward and poloidally upward towards X-point as the detachment degree increases, consistent with expected physical behavior. 
This capability highlights the robustness of DivControlNN in replicating key physical phenomena associated with divertor detachment, demonstrating its potential as a reliable tool for predicting and hence controlling divertor plasma behavior in the experiments.


\section{Deployment of DivControlNN and the preliminary results}\label{sec:app}
To evaluate the feasibility of machine-learning-based detachment control, a prototype detachment control system utilizing a single-model implementation of DivControlNN was deployed in the KSTAR Plasma Control System (PCS) during the December 2024 experimental campaign. This implementation aimed to assess the model’s real-time performance under experimental conditions, despite differences in diagnostic availability, impurity species, and device configuration compared to the training dataset. The following subsection describes the deployment process, key observations, and areas for further improvement.

There were a few differences in the assumptions and training dataset of DivControlNN in comparison to its experimental realization at KSTAR. As outlined in Sec.~\ref{subsec:modelio}, DivControlNN utilizes several input parameters, including plasma current ($I_p$), core electron density ($n_\mathrm{e}$), input power ($P_\mathrm{inj}$), impurity fraction ($f_Z$), and a diffusion profile scaling factor ($f_D$). In the KSTAR experiments though, only $I_p$ and $n_\mathrm{e}$ measurements were available in real-time diagnostics. Real-time $P_\mathrm{inj}$ was only available for certain neutral beam injection (NBI) sources, and thus, a pre-programmed input power waveform for other sources was added. $f_Z$ was calculated in real-time using a gas model that accounted for puffed gas ratios, decay rates, and wall loading, but this calculation malfunctioned during the experiments, and the model always received zero impurity fraction input. $f_D$ was set to a constant value of 1.0 based on prior estimates. Another key difference as mentioned in Sec.~\ref{subsec:database} was the device condition and impurity species. The UEDGE simulations used for training DivControlNN were based on KSTAR's earlier carbon divertor campaigns, assuming carbon as the sole impurity, while the latest KSTAR campaign used a tungsten divertor and the experiment used nitrogen as the impurity species for detachment control.

\begin{figure}[h]
    \centering
    \includegraphics[width=\linewidth]{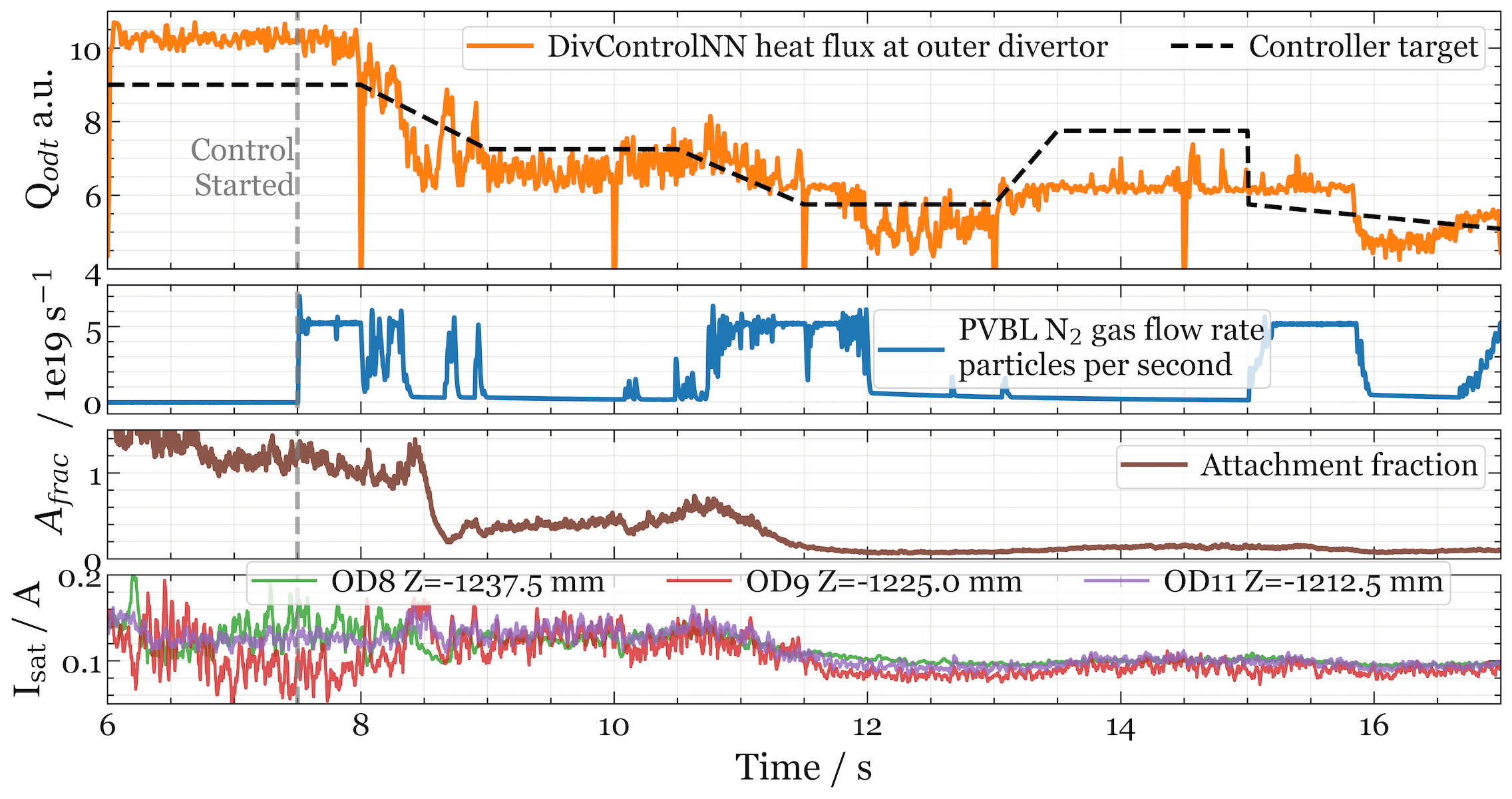}
    \caption{Demonstration of successful detachment control using the surrogate model, DivControlNN, on KSTAR shot \#36161. The black dashed line represents a prescribed control target sequence, while the \rev{orange} line indicates the real-time surrogate model prediction, which is used for real-time feedback control. }
    \label{fig:experiment}
\end{figure}

Despite these limitations, the preliminary results of using DivControlNN for real-time feedback detachment control exceeded expectations, highlighting the potential of this surrogate model for enabling advanced plasma control, even under non-ideal experimental conditions. 
\rev{In this experiment, a Proportional–Integral (PI) controller with anti-windup protection was employed, utilizing the DivControlNN-estimated peak heat flux at the outer divertor as its input (i.e., only one of DivControlNN's predictions was utilized) and the nitrogen gas command as its output~\cite{gputa2025detachment}.}
As shown in Figure~\ref{fig:experiment}, successful detachment control was achieved using DivControlNN’s prediction of the heat flux on the outer divertor target plate, $Q_\mathrm{odt}$, as the control target for KSTAR shot \#36161. 
The controller was initialized at 7.5 seconds, activating impurity (i.e., nitrogen in this case) gas puffing to gradually reduce the heat flux to the target value and push the outer divertor into a detached state at around 9 seconds. The detachment onset is evident in the attachment fraction, $A_\mathrm{frac}$, calculated via embedded Langmuir probes, which transitions from values near unity (e.g., at 7.5 seconds) to approximately 0.4 (e.g., at 9 seconds). This detached state was sustained until approximately 10.5 seconds, at which point the target heat flux was further reduced. Impurity seeding was reactivated, driving the system into a deeper detached state. This transition is reflected in the reduction of raw ion saturation current data from Langmuir probes near the strike point (OD8, OD9, and OD11), with the attachment fraction dropping below 0.1.
However, when the target heat flux was increased between 13 and 15 seconds and gas puffing stopped, the heat flux did not recover. This is likely due to excessive nitrogen injection and retention in the divertor region, which exceeded the controller's ability to restore the heat flux. 
\rev{The reason DivControlNN performed reasonably well with zero impurity fraction as an input in the early stage (e.g., from 8 to 12 seconds) is likely due to the effectiveness of other inputs in providing sufficient information for the model to generate approximately accurate heat flux predictions within certain limits. However, as the impurity concentration increased beyond a critical threshold, DivControlNN's output began to deviate from expectations substantially, particularly during continued nitrogen puffing. This mismatch resulted in degraded controller performance during the later stages of the test (e.g., after 15 seconds).}

We acknowledge that there is significant room to enhance the performance of DivControlNN further. For instance, training the surrogate model with a more accurate UEDGE database that incorporates the latest tungsten divertor configuration and non-carbon impurities could provide a better baseline for the model. Replacing the fixed fraction impurity model with a multi-charge-state impurity model in UEDGE simulations and consequently using the impurity seeding rate instead of the impurity fraction estimate in the model inputs could help reduce the uncertainty at the inputs of the model. Furthermore, thorough calibration could significantly improve the accuracy and overall effectiveness of DivControlNN. 
\rev{With the relevant divertor diagnostics collected in this KSTAR campaign (e.g., ion saturation current from Langmuir probes), an offline calibration and comparison between DivControlNN predictions and experimental diagnostics will be conducted, aiming to provide valuable insights into the model's performance and guide the design of improved controllers for future experiments.}
Nevertheless, these preliminary results are highly encouraging. Perhaps more importantly, the current model's robustness to variations in configuration and impurity species suggests a strong potential for transferability. In other words, the existing model can be leveraged for re-training on other devices and discharge scenarios using smaller simulation datasets and incorporating experimental corrections.

Incorporating the latest ensemble version of DivControlNN can also enhance the accuracy and robustness of detachment control by leveraging the strengths of multiple models to mitigate biases and improve prediction accuracy. Additionally, uncertainty quantification provides a direct measure of prediction confidence, enabling the control system to make adaptive adjustments in real-time to account for variations in plasma behavior and operational conditions.

\section{Summary}\label{sec:sum}

Fast boundary and divertor plasma models are essential for advancing magnetic fusion energy (MFE) research and development, particularly as the multi-scale, multi-physics complexity of these systems presents significant challenges for routine applications like real-time control and optimization in tokamaks. Recent advancements in machine learning have shown that boundary and divertor plasmas exhibit a low-dimensional latent space representation (LSR)~\cite{zhu2022data}, enabling the development of fast and reliable surrogate models. These models leverage LSR to provide manifold-consistent predictions across multi-modal datasets (e.g., images, spectra, profiles), making them highly versatile for general MFE applications.

In this work, we introduce DivControlNN, a novel machine-learning-based surrogate model designed for real-time detachment control in tokamaks. Trained on approximately 70,000 2D UEDGE simulations of KSTAR equilibria, DivControlNN delivers quasi-real-time predictions (i.e., around 0.2 milliseconds) of key plasma conditions with a relative error of less than 20\%. This represents a computational speed-up of over $10^8$ compared to traditional simulations, which often require hours to converge. The model self-consistently predicts critical boundary plasma conditions at multiple locations, including plasma profiles on divertor target plates (e.g., heat load, ion saturation current, and electron temperature), radiation information (e.g., overall and divertor radiation fractions, peak radiation intensity, and detachment front location), and upstream plasma conditions (e.g., electron density and temperature profiles at the outboard midplane). By leveraging its effective latent space mapping, DivControlNN ensures accurate and consistent predictions across diverse plasma scenarios. Notably, it is capable of predicting detachment onset - a critical operational regime where precise control is essential. Moreover, a prototype detachment controller based on DivControlNN was successfully deployed during the December 2024 KSTAR experimental campaign. Despite being trained on simulations assuming a carbon divertor and a fixed-fraction impurity model, the controller successfully achieved detachment control on a tungsten divertor configuration with nitrogen impurity seeding, highlighting the model's adaptability and transferability to different experimental conditions.

While DivControlNN has achieved significant milestones, there remains room for further development and broader applications. Future work will focus on expanding the training dataset to include tungsten divertor configurations and multi-charge-state impurity models, incorporating higher-fidelity data (e.g., from SOLPS with kinetic neutrals or even kinetic boundary transport models) \rev{through transfer learning~\cite{humbird2021cognitive,djordjevic2023transfer}}, calibrating and benchmarking model predictions against experimental results, \rev{and exploring effective methods for transferring the model to different experimental setups~\cite{ganin2016domain}.}

In summary, DivControlNN is a fast and reliable surrogate model that has demonstrated its potential to revolutionize divertor plasma control. Its ability to provide manifold-consistent predictions of multi-modal datasets makes it a powerful tool for real-time tokamak control and optimization. Beyond detachment control, the DivControlNN framework can be extended to other applications, such as discharge scenario development and multi-objective control, paving the way for advanced and integrated control systems in future fusion power plants.


\begin{acknowledgments}
This work was performed under the auspices of the U.S. Department of Energy (DOE) by Lawrence Livermore National Laboratory (LLNL) under Contract DE-AC52-07NA27344. This research used the resources of the National Energy Research Scientific Computing Center, a DOE Office of Science User Facility supported by the Office of Science of the U.S. Department of Energy under Contract No. DE-AC02-05CH11231. LLNL-JRNL-2003016
\end{acknowledgments}

\section*{Data Availability Statement}
The data that support the findings of this study are available from the corresponding author upon reasonable request.

\appendix
\section{Simulation datasets}\label{app:dataset}

The database of 2D UEDGE simulations is based on the KSTAR discharge \#22849 at 5.6 seconds. This was a lower-single-null H-mode discharge with a plasma current of 700 kA in the carbon divertor configuration. The simulation database is generated by varying five control parameters: the core boundary density, the power flowing through the core boundary into the simulation domain, the fraction of impurity ions relative to deuterium ions, the operational plasma current, and the scaling factor of the particle and thermal diffusivities (i.e., $D_\perp$, $\chi_i$, and $\chi_e$). The base profiles of $D_\perp$, $\chi_i$, and $\chi_e$ are derived by fitting the profiles used in SOLPS-ITER modeling of this KSTAR reference discharge, as shown in Fig.~\ref{fig:kstar_para_space}(a). These base profiles are divided into three regions, each characterized by distinct values (e.g., $D_\mathrm{core}$, $D_\mathrm{sep}$, and $D_\mathrm{SOL}$). The effective diffusivity profiles are constructed by connecting the core and the separatrix values using a hyperbolic tangent function and connecting the separatrix and the SOL values using an exponential function. To account for transport uncertainties in real experiments, a scaling factor $f_D$ is applied to the separatrix diffusivity values. For example, if $f_D > 1.0$ and $D_\mathrm{core}$ is lower than the scaled $D_\mathrm{sep}$, $D_\mathrm{core}$ is replaced with the scaled $D_\mathrm{sep}$ value. 

UEDGE uses a fully implicit Jacobian-free newton krylov method to evolve the boundary plasma transport equations. The convergence properties to obtain a steady state solution such as the number of steps required, are highly sensitive to the initial plasma conditions. To address this, an iterative data generation approach is employed. First, a UEDGE case in the middle of the targeted parameter range is performed and manually tuned to achieve a converged steady-state solution. This converged solution is then used as the initial guess for the first scan across all parameters. For subsequent cases that do not converge, a second scan is performed, during which the nearest converged solution is automatically selected as the initial guess. Often, after three iterations (or scans), the overall convergence ratio (i.e., the number of converged cases divided by the total number of launched cases) can exceed 50\%.

For the KSTAR simulation data used in this study, carbon impurities are assumed to be present in the system to account for radiation loss. A fixed-fraction model for carbon is employed to calculate the radiation loss. In this model, the carbon density and momentum equations are not solved. Instead, the ratio of the total number of carbon nuclei to the number of deuterium ions is treated as a user-specified control parameter, which is varied in the database. The radiation loss from carbon is calculated using a pre-defined table of carbon radiation loss rate coefficients. These coefficients are functions of the local electron density and temperature, assuming that the densities of each carbon charge state are balanced by ionization and recombination processes, based on collisional-radiative rates. Additionally, full cross-field drifts are enabled in the UEDGE simulations \rev{to better capture detachment onsets and plasma profiles at the divertor target plates~\cite{jaervinen2019impact}}.

Currently, Latin Hypercube Sampling (LHS) is used for parameter sampling when generating training dataset A and \rev{independent} validation dataset B. 
Datasets A and B both include plasma currents $I_\mathrm{p}=[600, 700, 800]$ kA. 
For dataset A, 30 sampling points are placed for $n_\mathrm{e,core}\in[1.5,7]\times10^{19}\mathrm{m}^{-3}$, 19 sampling points for $P_\mathrm{inj}\in[1,8]$ MW, 11 sampling points for $f_Z \in [0,0.04]$, and 9 sampling points for $f_D\in [0.6, 2]$ for each plasma current $I_\mathrm{p}$. 
The upper and lower bounds, as well as the sampling frequencies, are slightly adjusted for dataset B to ensure that it is independent of dataset A.
Specifically, for dataset B, 20 sampling points are placed for $n_\mathrm{e,core}\in[1.6,6.9]\times10^{19}\mathrm{m}^{-3}$, 8 sampling points for $P_\mathrm{inj}\in[1.2, 7.8]$ MW, 8 sampling points for $f_Z\in[1e-4, 0.039]$, and 7 sampling points for $f_D \in[0.65, 1.95]$ for each $I_\mathrm{p}$. 
This setup ensures that the parameter region of dataset B is close to, but independent of, dataset A. Table~\ref{tab:datasets} summarizes the numbers of converged UEDGE simulations and the corresponding convergence ratios for different plasma current setups in datasets A and B. As a result, datasets A and B exhibit similar distributions. For example, the detachment (labeled with $T_\mathrm{e,osp}<3.5$ eV) ratios are comparable: 51,459 detached cases are observed out of 70,659 total cases (72.83\%) in dataset A, while for dataset B, 11,701 detached cases are observed out of 16,293 total cases (71.82\%).




\begin{table}[h]
    \centering
    \begin{tabular}{|c|c|c|c|c|}
        \hline
        Dataset & $(n,f)_{I_p=600kA}$ & $(n,f)_{I_p=700kA}$ & $(n,f)_{I_p=800kA}$ & $(n,f)_\text{combined}$ \\ \hline
        A & 21,182 (58.84\%) & 25,365 (70.46\%) & 24,112 (66.98\%) & 70,659(65.43\%)\\ \hline
        B & 4,730 (52.80\%) & 5,745 (64.12\%) & 5,818 (64.94\%) & 16,293 (60.62\%) \\ \hline
    \end{tabular}
    \caption{Numbers of converged UEDGE simulations $n$ and the corresponding convergence ratios $f$ for different plasma current setups in datasets A and B.}
    \label{tab:datasets}
\end{table}



\section{Surrogate model performance}\label{app:perf}

\begin{figure}[h]
    \centering
    \includegraphics[width=\linewidth]{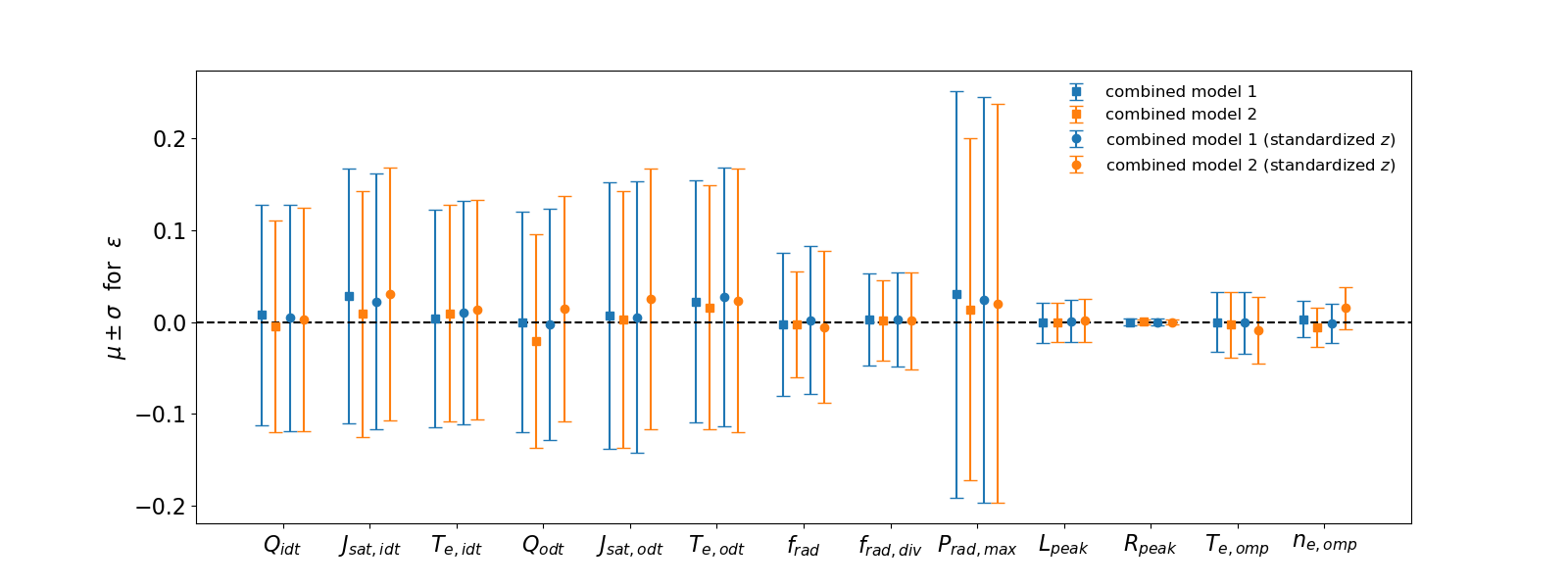}
    \caption{Performance of the surrogate model evaluated through the mean and standard deviation of prediction error distributions for combined ensemble models. All quantities are reported as relative errors, except for the peak radiation location (e.g., $L_{peak}$ and $R_{peak}$), which is reported in absolute error. }
    \label{fig:combined_perf}
\end{figure}

A detailed statistical analysis of the surrogate model's accuracy is presented in this appendix. The prediction accuracy of both individual models and combined ensemble models is evaluated for various plasma quantities, including divertor target profiles, radiation information, and upstream plasma conditions using the testing dataset B. The analysis is \rev{performed} in terms of the mean and standard deviation of prediction errors, \rev{with the results} summarized in \rev{Tables~\ref{tab:model_perf_stat_div}-\ref{tab:model_perf_stat_omp_regularized}, covering both native and standardized latent space representations (LSRs) mapping approaches}. For the combined models, combined model 1 incorporates all ten individual models, while combined model 2 consists of five selected individual models. Specifically, combined model 2 includes models \# 1, 5, 8, 9, and 10 based on native LSRs mapping and models \# 1, 6, 8, 9, and 10 based on standardized LSRs mapping. The statistical accuracy of these combined models is illustrated in Figure~\ref{fig:combined_perf}, which demonstrates that all four combined models exhibit similar performance. These statistics provide valuable insights into the reliability and variability of the surrogate model, as well as the influence of training parameters on prediction accuracy. This analysis further demonstrates the robustness and utility of DivControlNN for real-time plasma control applications. Additionally, the figures presented in Sections~\ref{sec:train} and~\ref{subsec:accuracy} are based on the results of model \# 9.

\begin{table}[h]
    \centering
    \begin{tabular}{|c|c|c|c|c|c|c|c|c|c|c|c||c|c|}
        \hline
         \multicolumn{2}{|c|}{model \#} & 1 & 2 & 3 & 4 & 5 & 6 & 7 & 8 & 9 & 10 & multi(10) & multi(5) \\ \hline
        \multirow{2}{*}{$Q_\mathrm{idt}$} & $\nu$ & 0.0290 & -0.0518 & -0.0102 & 0.0903 & 0.0214 & 0.0578 & -0.0211 & -0.0434 & -0.0217 & -0.0222 & 0.0077 & -0.0049 \\ \cline{2-14}
        & $\sigma$ & 0.1372 & 0.1518 & 0.1714 & 0.1446 & 0.1539 & 0.1449 & 0.1547 & 0.1307 & 0.1266 & 0.1258 & 0.1197 & 0.1149 \\ \hline
        \multirow{2}{*}{$J_\mathrm{sat,idt}$} & $\nu$ & 0.0328 & 0.0539 & 0.0215 & 0.0194 & 0.0267 & 0.1298 & -0.0573 & -0.0306 & 0.0133 & -0.0185 & 0.0284 & 0.0088 \\ \cline{2-14}
        & $\sigma$ & 0.1511 & 0.1601 & 0.1869 & 0.1524 & 0.1755 & 0.1639 & 0.1727 & 0.1537 & 0.1478 & 0.1479 & 0.1382 & 0.1335 \\ \hline
        \multirow{2}{*}{$T_\mathrm{e,idt}$} & $\nu$ & 0.0086 & -0.0525 & 0.0053 & -0.0184 & 0.0292 & 0.0066 & 0.0304 & -0.0184 & 0.0013 & 0.0123 & 0.0038 & 0.0094 \\ \cline{2-14}
        & $\sigma$ & 0.1339 & 0.1431 & 0.1644 & 0.1310 & 0.1478 & 0.1377 & 0.1494 & 0.1363 & 0.1310 & 0.1304 & 0.1178 & 0.1177 \\ \hline
        \multirow{2}{*}{$Q_\mathrm{odt}$} & $\nu$ & -0.0258 & -0.0932 & 0.0409 & 0.0014 & -0.0376 & 0.1307 & -0.0249 & -0.0554 & 0.0069 & -0.0077 & 0.0000 & -0.0208 \\ \cline{2-14}
        & $\sigma$ & 0.1366 & 0.1543 & 0.1657 & 0.1595 & 0.1516 & 0.1601 & 0.1468 & 0.1335 & 0.1356 & 0.1359 & 0.1200 & 0.1165 \\ \hline
        \multirow{2}{*}{$J_\mathrm{sat,odt}$} & $\nu$ & -0.0179 & -0.0593 & -0.0132 & -0.063 & 0.0204 & 0.1325 & -0.0228 & -0.0515 & 0.0132 & 0.0118 & 0.0071 & 0.0030 \\ \cline{2-14}
        & $\sigma$ & 0.1495 & 0.1623 & 0.1896 & 0.1786 & 0.1735 & 0.1712 & 0.1807 & 0.1576 & 0.1587 & 0.1578 & 0.1445 & 0.1397 \\ \hline
        \multirow{2}{*}{$T_\mathrm{e,odt}$} & $\nu$ & -0.0109 & -0.1188 & 0.1224 & -0.0005 & 0.0567 & 0.0603 & 0.0243 & -0.025 & 0.018 & 0.0104 & 0.0224 & 0.0159 \\ \cline{2-14}
        & $\sigma$ & 0.1612 & 0.1582 & 0.1796 & 0.1563 & 0.1735 & 0.1752 & 0.1589 & 0.1491 & 0.1465 & 0.1478 & 0.1318 & 0.1330 \\ \hline
    \end{tabular}
    \caption{The means and standard derivations of the surrogate model (via native LSR mapping) prediction errors for divertor target quantities.}
    \label{tab:model_perf_stat_div}
\end{table}

\begin{table}[h]
    \centering
    \begin{tabular}{|c|c|c|c|c|c|c|c|c|c|c|c||c|c|}
        \hline
         \multicolumn{2}{|c|}{model \#} & 1 & 2 & 3 & 4 & 5 & 6 & 7 & 8 & 9 & 10 & multi(10) & multi(5) \\ \hline
        \multirow{2}{*}{$Q_\mathrm{idt}$} & $\nu$ & 0.0405 & -0.0517 & -0.0227 & 0.0536 & 0.0118 & 0.0488 & -0.0022 & -0.0386 & -0.0230 & -0.0425 & 0.0044 & 0.0029 \\ \cline{2-14}
        & $\sigma$ & 0.1448 & 0.1512 & 0.1833 & 0.1632 & 0.1571 & 0.1550 & 0.1609 & 0.1384 & 0.1394 & 0.1365 & 0.1229 & 0.1212 \\ \hline
        \multirow{2}{*}{$J_\mathrm{sat,idt}$} & $\nu$ & 0.0493 & 0.0405 & 0.0129 & -0.0019 & 0.0014 & 0.1333 & -0.0402 & -0.0332 & -0.0069 & -0.0424 & 0.0223 & 0.0300 \\ \cline{2-14}
        & $\sigma$ & 0.1522 & 0.1634 & 0.1948 & 0.1704 & 0.1768 & 0.1664 & 0.1761 & 0.1562 & 0.1598 & 0.1577 & 0.1392 & 0.1374 \\ \hline
        \multirow{2}{*}{$T_\mathrm{e,idt}$} & $\nu$ & -0.0060 & -0.0396 & 0.0091 & -0.042 & 0.0410 & 0.0096 & 0.0359 & -0.0044 & 0.0110 & 0.0362 & 0.0103 & 0.0131 \\ \cline{2-14}
        & $\sigma$ & 0.1366 & 0.1468 & 0.1770 & 0.1481 & 0.1588 & 0.1436 & 0.1508 & 0.1389 & 0.1449 & 0.1409 & 0.1212 & 0.1193 \\ \hline
        \multirow{2}{*}{$Q_\mathrm{odt}$} & $\nu$ & -0.0138 & -0.0963 & 0.0340 & -0.0145 & -0.0698 & 0.1397 & -0.0098 & -0.047 & -0.0169 & -0.0327 & -0.0025 & 0.0143 \\ \cline{2-14}
        & $\sigma$ & 0.1544 & 0.1696 & 0.1791 & 0.1784 & 0.1615 & 0.1613 & 0.1573 & 0.1431 & 0.1485 & 0.1462 & 0.1253 & 0.1225 \\ \hline
        \multirow{2}{*}{$J_\mathrm{sat,odt}$} & $\nu$ & 0.0004 & -0.0672 & -0.0157 & -0.047 & -0.0032 & 0.1282 & -0.0094 & -0.0456 & -0.0032 & -0.0118 & 0.0054 & 0.0250 \\ \cline{2-14}
        & $\sigma$ & 0.1622 & 0.1729 & 0.1987 & 0.1947 & 0.1795 & 0.1764 & 0.1902 & 0.1646 & 0.1660 & 0.1666 & 0.1472 & 0.1420 \\ \hline
        \multirow{2}{*}{$T_\mathrm{e,odt}$} & $\nu$ & -0.0249 & -0.1193 & 0.1281 & -0.0061 & 0.0285 & 0.0890 & 0.0187 & -0.0095 & 0.0004 & 0.0136 & 0.0273 & 0.0230 \\ \cline{2-14}
        & $\sigma$ & 0.1809 & 0.1791 & 0.1908 & 0.1837 & 0.1928 & 0.1775 & 0.1734 & 0.1625 & 0.1670 & 0.1660 & 0.1411 & 0.1433 \\ \hline
    \end{tabular}
    \caption{The means and standard derivations of the surrogate model (via standardized LSR mapping) prediction errors for divertor target quantities.}
    \label{tab:model_perf_stat_div_regularized}
\end{table}

\begin{table}[h]
    \centering
    \begin{tabular}{|c|c|c|c|c|c|c|c|c|c|c|c||c|c|}
        \hline
         \multicolumn{2}{|c|}{model \#} & 1 & 2 & 3 & 4 & 5 & 6 & 7 & 8 & 9 & 10 & multi(10) & multi(5) \\ \hline
        \multirow{2}{*}{$f_\mathrm{rad}$} & $\nu$ & -0.0111 & -0.0112 & 0.0488 & -0.0297 & 0.0027 & -0.0249 & 0.0050 & -0.0071 & -0.0186 & 0.0236 & -0.0026 & -0.0024 \\ \cline{2-14}
        & $\sigma$ & 0.0548 & 0.1196 & 0.0833 & 0.1314 & 0.0510 & 0.1283 & 0.1281 & 0.1304 & 0.0545 & 0.1229 & 0.0778 & 0.0577 \\ \hline
        \multirow{2}{*}{$f_\mathrm{rad,div}$} & $\nu$ & -0.0053 & 0.0076 & -0.0116 & 0.0148 & 0.0082 & 0.0016 & 0.0056 & 0.0060 & -0.0050 & 0.0040 & 0.0026 & 0.0016 \\ \cline{2-14}
        & $\sigma$ & 0.0396 & 0.0675 & 0.0420 & 0.0665 & 0.0410 & 0.0663 & 0.0662 & 0.0652 & 0.0363 & 0.0731 & 0.0503 & 0.0440 \\ \hline
        \multirow{2}{*}{$P_\mathrm{rad,max}$} & $\nu$ & -0.0107 & 0.0472 & 0.0233 & 0.0034 & -0.0172 & 0.0701 & 0.0370 & 0.0509 & -0.0326 & 0.0454 & 0.0301 & 0.0136 \\ \cline{2-14}
        & $\sigma$ & 0.1752 & 0.3132 & 0.1998 & 0.3054 & 0.1811 & 0.3024 & 0.3106 & 0.3043 & 0.1568 & 0.3070 & 0.2209 & 0.1859 \\ \hline
        \multirow{2}{*}{$L_\mathrm{peak}$} & $\nu$ & -0.0024 & 0.0034 & 0.0011 & -0.0065 & 0.0012 & -0.0024 & -0.0014 & -0.0017 & -0.0025 & 0.0035 & -0.0008& -0.0004 \\ \cline{2-14}
        & $\sigma$ & 0.0220 & 0.0231 & 0.0228 & 0.0229 & 0.0235 & 0.0230 & 0.0233 & 0.0236 & 0.0222 & 0.0254 & 0.0215 & 0.0216 \\ \hline
        \multirow{2}{*}{$R_\mathrm{peak}$} & $\nu$ & 0.0002 & 0.0006 & 0.0001 & -0.0010 & 0.0003 & 0.0000 & 0.0000 & 0.0001 & -0.0003 & 0.0004 & 0.0000 & 0.0001 \\ \cline{2-14}
        & $\sigma$ & 0.0026 & 0.0038 & 0.0039 & 0.0038 & 0.0038 & 0.0039 & 0.0038 & 0.0038 & 0.0025 & 0.0036 & 0.0033 & 0.0030 \\ \hline
    \end{tabular}
    \caption{The means and standard derivations of the surrogate model (via native LSR mapping) prediction errors for divertor region radiation information.}
    \label{tab:model_perf_stat_rad}
\end{table}

\begin{table}[h]
    \centering
    \begin{tabular}{|c|c|c|c|c|c|c|c|c|c|c|c||c|c|}
        \hline
         \multicolumn{2}{|c|}{model \#} & 1 & 2 & 3 & 4 & 5 & 6 & 7 & 8 & 9 & 10 & multi(10) & multi(5) \\ \hline
        \multirow{2}{*}{$f_\mathrm{rad}$} & $\nu$ & -0.0058 & 0.0115 & 0.0520 & -0.0301 & 0.0216 & -0.0446 & -0.0056 & 0.0033 & 0.0161 & 0.0044 & 0.0019 & -0.0053 \\ \cline{2-14}
        & $\sigma$ & 0.0689 & 0.1537 & 0.1121 & 0.1372 & 0.0756 & 0.1263 & 0.1337 & 0.1412 & 0.0849 & 0.1601 & 0.0806 & 0.0825 \\ \hline
        \multirow{2}{*}{$f_\mathrm{rad,div}$} & $\nu$ & 0.0046 & 0.0067 & -0.0156 & 0.0142 & 0.0047 & 0.0010 & 0.0100 & -0.0011 & 0.0027 & -0.0006 & 0.0027 & 0.0014 \\ \cline{2-14}
        & $\sigma$ & 0.0440 & 0.0683 & 0.0454 & 0.0693 & 0.0400 & 0.0680 & 0.0646 & 0.0640 & 0.0408 & 0.0767 & 0.0510 & 0.0525 \\ \hline
        \multirow{2}{*}{$P_\mathrm{rad,max}$} & $\nu$ & 0.0373 & 0.0286 & 0.0245 & 0.0017 & 0.0250 & 0.0311 & 0.0568 & 0.0239 & 0.0086 & -0.0222 & 0.0242 & 0.0203 \\ \cline{2-14}
        & $\sigma$ & 0.1970 & 0.3179 & 0.2339 & 0.3187 & 0.2029 & 0.3157 & 0.3116 & 0.3094 & 0.2006 & 0.2976 & 0.2207 & 0.2166 \\ \hline
        \multirow{2}{*}{$L_\mathrm{peak}$} & $\nu$ & -0.0026 & 0.0055 & -0.0027 & -0.0063 & 0.0050 & -0.0055 & -0.0025 & -0.0045 & 0.0008 & 0.0025 & -0.0010 & 0.0019 \\ \cline{2-14}
        & $\sigma$ & 0.0259 & 0.0245 & 0.0266 & 0.0279 & 0.0268 & 0.0262 & 0.0244 & 0.0257 & 0.0260 & 0.0289 & 0.0229 & 0.0234 \\ \hline
        \multirow{2}{*}{$R_\mathrm{peak}$} & $\nu$ & 0.0001 & 0.0008 & 0.0003 & -0.0010 & 0.0006 & -0.0003 & 0.0000 & -0.0030 & 0.0000 & 0.0002 & 0.0000 & -0.0001 \\ \cline{2-14}
        & $\sigma$ & 0.0028 & 0.0038 & 0.0041 & 0.0040 & 0.0039 & 0.0038 & 0.0040 & 0.0038 & 0.0028 & 0.0038 & 0.0033 & 0.0030 \\ \hline
    \end{tabular}
    \caption{The means and standard derivations of the surrogate model (via standardized LSR mapping) prediction errors for divertor region radiation information.}
    \label{tab:model_perf_stat_rad_regularized}
\end{table}

\begin{table}[h]
    \centering
    \begin{tabular}{|c|c|c|c|c|c|c|c|c|c|c|c||c|c|}
        \hline
         \multicolumn{2}{|c|}{model \#} & 1 & 2 & 3 & 4 & 5 & 6 & 7 & 8 & 9 & 10 & multi(10) & multi(5) \\ \hline
        \multirow{2}{*}{$T_\mathrm{e,omp}$} & $\nu$ & -0.0002 & -0.0468 & 0.0204 & 0.0376 & 0.0286 & 0.0007 & 0.0020 & -0.0070 & -0.0139 & -0.0223 & -0.0001 & -0.0029 \\ \cline{2-14}
        & $\sigma$ & 0.0484 & 0.0619 & 0.0648 & 0.0513 & 0.0357 & 0.0371 & 0.0534 & 0.0509 & 0.0442 & 0.0546 & 0.0325 & 0.0360 \\ \hline
        \multirow{2}{*}{$n_\mathrm{e,omp}$} & $\nu$ & 0.0209 & 0.0437 & -0.0418 & 0.0029 & -0.0498 & 0.0688 & -0.0173 & 0.0066 & 0.0043 & -0.0087 & 0.0029 & -0.0054 \\ \cline{2-14}
        & $\sigma$ & 0.0368 & 0.0415 & 0.0486 & 0.0516 & 0.0512 & 0.0321 & 0.0261 & 0.0271 & 0.0312 & 0.0368 & 0.0196 & 0.0214 \\ \hline
    \end{tabular}
    \caption{The means and standard derivations of the surrogate model (via native LSR mapping) prediction errors for outboard midplane plasma quantities.}
    \label{tab:model_perf_stat_omp}
\end{table}

\begin{table}[h]
    \centering
    \begin{tabular}{|c|c|c|c|c|c|c|c|c|c|c|c||c|c|}
        \hline
         \multicolumn{2}{|c|}{model \#} & 1 & 2 & 3 & 4 & 5 & 6 & 7 & 8 & 9 & 10 & multi(10) & multi(5) \\ \hline
        \multirow{2}{*}{$T_\mathrm{e,omp}$} & $\nu$ & 0.0039 & -0.049 & 0.0174 & 0.0300 & 0.0273 & -0.0046 & 0.0122 & -0.0052 & -0.0121 & -0.0274 & -0.0007 & -0.0091 \\ \cline{2-14}
        & $\sigma$ & 0.0524 & 0.0646 & 0.0755 & 0.0645 & 0.0419 & 0.0405 & 0.0547 & 0.0546 & 0.0500 & 0.0583 & 0.0325 & 0.0362 \\ \hline
        \multirow{2}{*}{$n_\mathrm{e,omp}$} & $\nu$ & 0.0341 & 0.0418 & -0.0477 & -0.0073 & -0.0572 & 0.0689 & -0.0181 & 0.0065 & -0.0051 & -0.0295 & -0.0014 & -0.0150 \\ \cline{2-14}
        & $\sigma$ & 0.0419 & 0.0538 & 0.0534 & 0.0604 & 0.0549 & 0.0373 & 0.0378 & 0.0314 & 0.0444 & 0.0426 & 0.0209 & 0.0225 \\ \hline
    \end{tabular}
    \caption{The means and standard derivations of surrogate model (via standardized LSR mapping) prediction errors for outboard midplane plasma quantities.}
    \label{tab:model_perf_stat_omp_regularized}
\end{table}

\nocite{*}
\bibliography{sdetach_2d_r1}

\end{document}